\documentclass[aps,prd,reprint,superscriptaddress,nofootinbib]{revtex4-2}
\usepackage{amsmath,amssymb,amsfonts}
\usepackage{graphicx}
\usepackage[caption=false]{subfig}
\usepackage{bm}
\usepackage[colorlinks=true, pdfstartview=FitV, linkcolor=blue, citecolor=red, urlcolor=magenta, breaklinks=true]{hyperref}

\begin{document}

\title{Near-boundary and near-horizon momentum--Krylov response in noncommutative-inspired BTZ spacetime}

\author{F. A. Brito}
\email{fabrito@df.ufcg.edu.br}
\affiliation{Departamento de F\'\i sica, Universidade Federal de Campina Grande, Caixa Postal 10071, 58429-900 Campina Grande, Para\'\i ba, Brazil}

\author{F. G. Costa}
\email{francisco.geraldo@ifpb.edu.br}
\affiliation{Instituto Federal de Educa\c{c}\~ao, Ci\^encia e Tecnologia da Para\'\i ba (IFPB), Campus Campina Grande, Rua Tranquilino Coelho Lemos 671, Jardim Dinam\'erica I, Campina Grande, Para\'\i ba, Brazil}

\author{J. A. V. Campos}
\email{a.campos@uaf.ufcg.edu.br}
\affiliation{Departamento de F\'\i sica, Universidade Federal de Campina Grande, Caixa Postal 10071, 58429-900 Campina Grande, Para\'\i ba, Brazil}

\begin{abstract}
We compute the proper radial momentum of a massive probe falling into a Lorentzian-smeared, noncommutative-inspired BTZ black hole at fixed asymptotic mass, and use it as the growth rate of momentum--Krylov complexity. In the BTZ limit the accumulated complexity is $\mathcal{K}_0 = X(\cosh\tau - 1)$. With smearing, the initial growth coefficient is $F'_\alpha(X)/4$, while the late-time exponent is the surface gravity $F'_\alpha(x_+)/2$. The two depend differently on the smearing parameter, so the deformed response is not a rescaled BTZ curve; we show this analytically and confirm it numerically. For a distant release point the early response approaches BTZ, whereas the late-time rate does not. A BTZ black hole with the same Hawking temperature reproduces the response to within about ten percent, and the remaining difference is set by the early/late mismatch. A Gaussian mass profile gives the same picture. The horizon radius and the surface gravity vanish at the critical smearing with exponents $1/2$ and $3/2$. At fixed mass the boundary stress tensor coincides with that of BTZ, but the $z^3$ Fefferman--Graham coefficient and the lengths of boundary-anchored geodesics do not.
\end{abstract}

\maketitle

\section{Introduction}

The BTZ black hole provides one of the cleanest laboratories for black-hole physics and AdS$_3$/CFT$_2$ holography \cite{Carlip:1995qv,AdS,BTZ}. Its simplicity has also made it useful for studying deformations of three-dimensional gravity, including scalar and dilaton sectors \cite{Chan:1995wj,Martinez:1996gn,Chan:1996rd} and noncommutative-inspired geometries \cite{Banados:2001,Kim:2008,Chang-Young:2009,Jahangir:2011,Liang:2012}. In the latter approach, a point source is replaced by a smooth distribution, following the general noncommutative-inspired black-hole program \cite{Nicolini:2005vd,Nicolini:2008aj,Nozari:2008rc}. Lorentzian-smeared BTZ backgrounds have subsequently been used in thermodynamics, scattering, and related gravitational phenomena \cite{Anacleto:2014cga,Anacleto:2019tdj,Rahaman:2013gw,Sadeghi:2015nzp,Anacleto:2015kca,Kawamoto:2017ufe}.

Quantum complexity offers another probe of gravitational dynamics. Circuit complexity, spread complexity, and operator Krylov complexity are different constructions, and comparing them requires specifying the state, the operator, and the Krylov space \cite{Nandy:2024evd,Baiguera:2025dkc,Rabinovici:2025otw}. The holographic CV and CA proposals \cite{Douglas:2014,Adam:2016,Brown:2016,Chapman:2017} are also distinct from the proper-momentum prescription used here.

Caputa \emph{et al.}\ related the rate of spread complexity of locally excited states in two-dimensional CFTs to the proper radial momentum of a particle in the AdS$_3$ dual \cite{Caputa:2026}. Fan proposed a momentum--Krylov correspondence and realized it in BTZ and other AdS black holes \cite{Fan:2024iop}. The relation has been analysed further in Refs.~\cite{He:2024,Alfinito:2026} and extended to more general holographic theories \cite{Fatemiabhari:2026}. Recent BTZ studies also show that spread and operator Krylov complexity can behave differently \cite{Bhattacharya:2026quadratic,Bhattacharyya:2026interior}. We therefore refer to
\begin{equation}
 \dot{\mathcal C}=|p_\rho|,
 \label{eq:maincorrespondence}
\end{equation}
as the \emph{momentum--Krylov prescription}, where $p_\rho$ is the momentum conjugate to proper radial distance.

Black-hole horizons are also the natural setting for the chaos bound $\lambda_L\leq 2\pi k_B T/\hbar$ of Maldacena, Shenker, and Stanford (MSS), formulated for the exponential regime of thermal out-of-time-order correlators \cite{Maldacena:2015waa}. Near the horizon the proper momentum grows at a rate set by the surface gravity, so its exponent can be compared with this thermal scale. We compare scales only: the probe exponent is not computed from an out-of-time-order correlator, and Krylov growth and chaos indicators need not coincide \cite{Avdoshkin:2022xuw,Hashimoto:2023swv}.

Our main question is how the smearing parameter changes the full time-dependent response at fixed asymptotic mass, beyond the near-horizon exponent. We use the unexpanded Lorentzian mass profile throughout, so the probe moves in a single smooth background. We obtain the outer horizon in closed form, solve the BTZ case exactly, derive the trajectory and proper momentum for general smearing, and determine the critical scaling of the horizon and the surface gravity. Independently of the complexity interpretation, we also compute boundary-anchored spatial geodesics, which distinguish the smeared geometry although its boundary stress tensor equals that of BTZ.

The early-time and late-time regimes are governed by different coefficients, and the comparison between them is the core of the paper. Over finite times the resulting change of shape is at the percent level, so we separate the analytic statement from its numerical size and compare with BTZ black holes of the same mass and of the same Hawking temperature. Here ``near-boundary'' and ``near-horizon'' refer to the release region and the horizon region of the probe; they are unrelated to UV/IR mixing in noncommutative field theory. A Gaussian mass profile serves as a second example.

Holographically, one may ask whether the smeared geometry corresponds to a thermofield-double (TFD) state of the undeformed CFT at a shifted temperature, or to a state with additional operator content. Operator-deformed thermofield geometries in AdS$_3$ provide a precedent for the latter \cite{Bak:2017}. We discuss this question but do not settle it: the effective source and the near-horizon momentum law do not fix a boundary Hamiltonian, a density matrix, or a set of Lanczos coefficients.

The paper is organized as follows. Section~\ref{sec:geometry} introduces the Lorentzian-smeared BTZ geometry and its outer horizon. Section~\ref{sec:holographic} discusses the Fefferman--Graham expansion, the boundary stress tensor, and the holographic interpretation. Section~\ref{sec:formalism} reviews the proper radial momentum prescription. Section~\ref{sec:application} develops the dimensionless dynamics, the exact BTZ solution, and numerical checks. Section~\ref{sec:thetaresponse} computes the finite-time response and proves that it cannot be reduced to a two-parameter rescaling of the BTZ curve. Section~\ref{sec:nonlocal} presents the geodesic lengths and the critical scaling. Section~\ref{sec:discussion} contains our conclusions.

\section{Lorentzian-smeared BTZ geometry}\label{sec:geometry}

We consider the Lorentzian mass density \cite{Liang:2012,Anacleto:2019tdj}
\begin{equation}
 \rho_\theta(r)=\frac{M\sqrt\theta}{2\pi(r^2+\theta)^{3/2}},
 \label{eq:rho}
\end{equation}
where $\theta$ has dimensions of length squared. The mass inside radius $r$ is
\begin{equation}
 {\cal M}_\theta(r)=\int_0^r2\pi r'\rho_\theta(r')\,dr'
 =M\left(1-\frac{\sqrt\theta}{\sqrt{r^2+\theta}}\right).
 \label{eq:Mexact}
\end{equation}
For the nonrotating geometry we take
\begin{equation}
 ds^2=-f_\theta(r)dt^2+\frac{dr^2}{f_\theta(r)}+r^2d\phi^2,
 \label{eq:metric}
\end{equation}
with
\begin{equation}
 f_\theta(r)=\frac{r^2}{l^2}
 -M\left(1-\frac{\sqrt\theta}{\sqrt{r^2+\theta}}\right).
 \label{eq:fexact}
\end{equation}
The BTZ metric is recovered as $\theta\to0$ at fixed $r>0$. Throughout the paper Eq.~\eqref{eq:fexact} is used without a large-radius expansion.

For completeness, the geometry is consistent with the anisotropic effective source associated with Eq.~\eqref{eq:rho}. In units $8G=1$, Einstein's equations with $\Lambda=-1/l^2$ give
\begin{equation}
 T^t{}_t=T^r{}_r=-\rho_\theta,
 \qquad
 T^\phi{}_\phi=-\left(\rho_\theta+r\rho_\theta'\right),
 \label{eq:sourcepressures}
\end{equation}
or $p_r=-\rho_\theta$ and $p_\phi=-\rho_\theta-r\rho_\theta'$. This is an on-shell reconstruction of the stress tensor required by the chosen metric, rather than a derivation of that metric from a specified fixed-coupling matter action. The subsequent probe equations depend on the metric and do not require a microscopic identification of this source.

It is convenient to introduce the commutative BTZ horizon scale $r_h=l\sqrt M$ and
\begin{equation}
 x=\frac{r}{r_h},\qquad
 \alpha=\frac{\sqrt\theta}{r_h}.
 \label{eq:dimensionlessgeom}
\end{equation}
Then
\begin{equation}
 \frac{f_\theta(r)}{M}=F_\alpha(x)
 =x^2-1+\frac{\alpha}{\sqrt{x^2+\alpha^2}}.
 \label{eq:Falpha}
\end{equation}
The finite-radius outer Killing horizon is the positive zero of $F_\alpha$. It is obtained exactly as
\begin{equation}
 x_+^2=1-\frac{\alpha^2}{2}
 -\frac{\alpha}{2}\sqrt{\alpha^2+4},
 \label{eq:xplusexact}
\end{equation}
which is positive for
\begin{equation}
 0\leq\alpha<\alpha_c,\qquad \alpha_c=\frac{1}{\sqrt2}.
 \label{eq:alphacrit}
\end{equation}
For $\alpha>0$, $x=0$ is also a degenerate zero of $F_\alpha$. Since its areal radius vanishes, we keep it distinct from the finite-radius outer Killing horizon relevant for the exterior infalling probe. At $\alpha=\alpha_c$, the finite-radius root approaches the core root. No global interpretation of the zero-radius root is required for the exterior dynamics studied here.

\begin{figure}[htbp]
 \centering
  \includegraphics[width=\columnwidth]{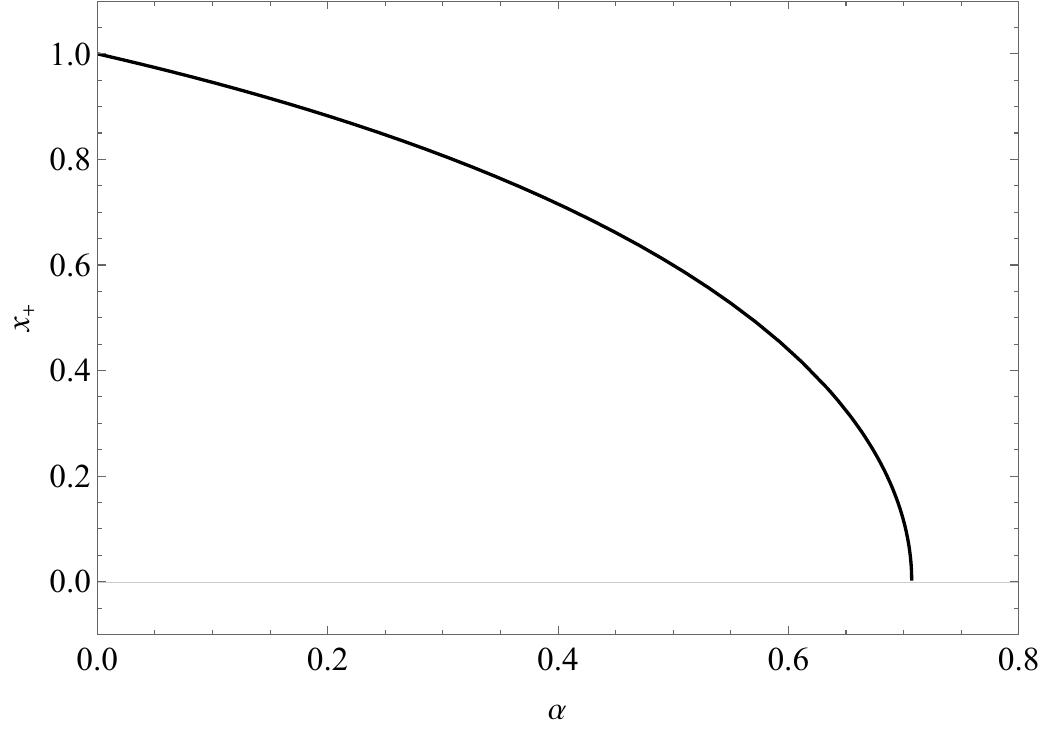}
  \caption{\footnotesize{Finite-radius outer Killing horizon $x_+$ as a function of the Lorentzian smearing parameter $\alpha$. The BTZ value $x_+=1$ is recovered at $\alpha=0$, while $x_+\to0$ as $\alpha\to1/\sqrt2$ from below.}}
  \label{fig:horizonalpha}
 \end{figure}

Figure~\ref{fig:horizonalpha} shows the monotonic inward displacement of the outer horizon with increasing smearing.

\section{Holographic status of the smeared background}\label{sec:holographic}

\subsection{Thermofield interpretation and geometric data}

For an eternal ordinary BTZ black hole, the standard thermofield-double construction is \begin{equation}\label{eq:tfd}\lvert\mathrm{TFD}(\beta)\rangle=Z(\beta)^{-1/2}\sum_n e^{-\beta E_n/2}\lvert E_n\rangle_L\lvert E_n\rangle_R.\end{equation} The Lorentzian smearing changes the Hawking temperature; however, it is not evident whether the same CFT Hamiltonian and the same thermal state construction remain sufficient, or whether the dual requires an excited or operator-deformed thermofield state. The operator-deformed BTZ constructions of Ref.~\cite{Bak:2017} illustrate that backreaction may encode nonthermal state data without necessarily changing the boundary Hamiltonian.

The bulk geometry distinguishes these interpretations at the level of local invariants. Writing $q=\sqrt{\theta}$, its scalar curvature is
\begin{equation}
 R=-f_\theta''-\frac{2f_\theta'}{r}
 =-\frac{6}{l^2}+\frac{3Mq^3}{(r^2+q^2)^{5/2}}.
 \label{eq:ricci_holo}
\end{equation}
Unlike vacuum BTZ, the smeared solution is not locally of constant curvature for $q>0$. Thus a temperature or mass redefinition cannot make the two bulk metrics locally equivalent. This observation does \emph{not}, by itself, distinguish a deformed CFT Hamiltonian from a nontrivial state of an unchanged CFT: that distinction requires the matter action and its boundary conditions.

The asymptotic expansion
\begin{equation}
 f_\theta(r)=\frac{r^2}{l^2}-M+\frac{Mq}{r}-\frac{Mq^3}{2r^3}+O(r^{-5})
 \label{eq:asymptotic_holo}
\end{equation}
retains the leading BTZ falloffs. Consequently, Brown--Henneaux-type asymptotic conditions are plausible for the metric sector \cite{BrownHenneaux:1986}, but the existence and nature of the complete dual theory depend on the asymptotic behavior and variational principle of the supporting matter fields. In particular, neither an ordinary TFD at $T_H$ nor an operator-deformed TFD has been derived from Eq.~\eqref{eq:fexact}.

\subsection{Fefferman--Graham data and the boundary stress tensor}\label{sec:fg}

A natural way to address the thermofield question is to ask whether the Lorentzian smearing is visible in the holographic energy--momentum tensor. For the metric in Eq.~\eqref{eq:fexact}, the Fefferman--Graham (FG) expansion shows that the first explicit dependence on $q=\sqrt\theta$ occurs beyond the coefficient $g_{(2)ij}$ that determines the standard gravitational stress tensor in AdS$_3$. This behavior motivates a comparison of the asymptotic stress tensors before attributing the deformation to a change of boundary Hamiltonian or to a particular thermofield state.

To see this, introduce a dimensionless FG coordinate $z$ and a boundary spatial coordinate $x=l\phi$. The radial gauge condition is
\begin{equation}
 \frac{dr^2}{f_\theta(r)}=\frac{l^2}{z^2}dz^2,
 \qquad \frac{dr}{dz}=-\frac{l}{z}\sqrt{f_\theta(r)}.
 \label{eq:fg_gauge}
\end{equation}
With the boundary normalization $r\sim l/z$, the asymptotic solution is
\begin{equation}
 r(z)=\frac{l}{z}+\frac{Ml}{4}z-\frac{Mq}{6}z^2+O(z^3).
 \label{eq:fg_radial}
\end{equation}
Writing
\begin{equation}
\begin{split}
 ds^2={}&\frac{l^2dz^2}{z^2}\\
 &+\frac{1}{z^2}\left[g_{(0)ij}+z^2g_{(2)ij}+z^3g_{(3)ij}+O(z^4)\right]dx^idx^j,
\end{split}
 \label{eq:fg_expansion}
\end{equation}
where $x^i=(t,x)$, we obtain
\begin{align}
 g_{(0)ij}&=\begin{pmatrix}-1&0\\0&1\end{pmatrix},
 &g_{(2)ij}&=\frac{M}{2}\begin{pmatrix}1&0\\0&1\end{pmatrix},\nonumber\\
 g_{(3)ij}&=-\frac{Mq}{3l}\begin{pmatrix}2&0\\0&1\end{pmatrix}.
 \label{eq:fg_coefficients}
\end{align}
Hence $\partial_q g_{(2)ij}=0$, whereas $g_{(3)ij}$ records the Lorentzian smearing. The $z^3$ term is permitted here because the bulk is supported by an effective matter source; the pure Einstein-vacuum FG recursion should not be imposed on this nonvacuum geometry.

The standard AdS$_3$ stress-tensor formula consequently yields the same gravitational boundary stress tensor for the ordinary and Lorentzian-smeared BTZ geometries at fixed asymptotic mass $M$, even though their horizon temperatures differ. The smearing is therefore not captured by this particular one-point observable.

For a flat boundary metric, $\operatorname{Tr}_{g_{(0)}}g_{(2)}=0$, and the conventional renormalized Einstein-gravity expression \cite{deHaro:2000} gives
\begin{equation}
\begin{split}
 \langle T_{ij}\rangle_{\rm grav}&=
 \frac{1}{8\pi G l}\left(g_{(2)ij}-g_{(0)ij}
 \operatorname{Tr}_{g_{(0)}}g_{(2)}\right)\\
 &=\frac{M}{16\pi G l}\begin{pmatrix}1&0\\0&1\end{pmatrix}.
\end{split}
 \label{eq:holo_tem}
\end{equation}
The components in Eq.~\eqref{eq:holo_tem} refer to the physical-length spatial coordinate $x=l\phi$ and the FG boundary representative as written in Eq.~\eqref{eq:fg_expansion}; rescaling boundary coordinates or the boundary metric changes their component normalization. In this convention the stress tensor is traceless. Integrating its energy density over the boundary circle of circumference $2\pi l$ gives $E_{\rm grav}=M/(8G)$, consistently with the gravitational charge used below (up to the AdS reference energy and possible matter contributions). This equality is a statement about the \emph{standard gravitational contribution} to the renormalized stress tensor: the full holographic one-point function would also require the matter action, boundary conditions, and counterterms.

Equality of $\langle T_{ij}\rangle$ at fixed $M$ does not establish equality of boundary states, equality of their temperatures, or equivalence of their microscopic CFT descriptions. The Hawking temperature of the smeared solution depends on $q$, while the displayed stress tensor does not. A conventional canonical thermal state of an otherwise unchanged CFT cannot be inferred from these data alone; additional state or source information may be required. Quantitatively, in the high-temperature, large-central-charge regime of the ordinary BTZ/CFT$_2$ canonical ensemble, $E_{\rm th}(T)\propto T^2$. Comparing at the same $M$ gives
\begin{equation}
 \frac{E_{\rm th}(T_H)}{E_{\rm th}(T_{\rm BTZ})}=\hat\kappa^2,
 \qquad \frac{S_{\rm BH}(q)}{S_{\rm BTZ}}=x_+.
 \label{eq:thermal_comparison}
\end{equation}
For example, these ratios are approximately $(0.38,0.81)$ at $\alpha=0.3$ and $(0.081,0.60)$ at $\alpha=0.5$. They measure the mismatch between the fixed asymptotic gravitational energy and the energy of an \emph{ordinary} thermal state of the same CFT at the lower Hawking temperature. They do not establish a new nonthermal charge, nor a specific deformed TFD: matter boundary terms, the correct ensemble, and the dual state are still undetermined. The coefficient $g_{(3)}$ is geometric evidence of subleading bulk data, but is not, by itself, an identified expectation value of a specific boundary operator. Nonlocal correlators and state-sensitive probes could discriminate the geometries, whereas the present paper shows this sensitivity only for the \emph{geometric} proper-momentum prescription.

\subsection{Source consistency and the scope of a dynamical completion}

The anisotropic source obeys $\rho+p_r=0$ and $\rho+p_\phi=-r\rho'\geq0$, so its null and weak energy conditions hold for $M,q>0$ with the source normalization of Eq.~\eqref{eq:rho}. The dominant energy condition need not hold: $p_\phi/\rho=(2r^2-q^2)/(r^2+q^2)>1$ for $r^2>2q^2$. A minimally coupled static radial canonical scalar cannot reproduce $T^t{}_t=T^r{}_r$ with a nonconstant profile outside the horizon.

One possible effective parametrization of the source uses nonlinear electrodynamics (NED), written in Hamiltonian variables $P_{\mu\nu}=\mathcal L_{\mathcal F}F_{\mu\nu}$ and $\mathcal P=P_{\mu\nu}P^{\mu\nu}$. For a circular electric solution, the field equation gives $\mathcal P\propto-Q^2/r^2$, which is monotonic for $r>0$. A Hamiltonian density with the schematic dependence
\begin{equation}
 \mathcal H(\mathcal P)=a\left(q^2-\frac{2Q_0^2}{\mathcal P}\right)^{-3/2},\qquad \mathcal P<0,
 \label{eq:ned_hamiltonian}
\end{equation}
has the radial functional dependence needed to reproduce the Lorentzian profile after parameter matching and stress-tensor normalization. This observation is not a construction of a single fixed-coupling action generating the full two-parameter $(M,q)$ family. Its inverse Legendre map to a single-valued $\mathcal L(\mathcal F)$ is not globally invertible: the associated electric invariant returns to zero both at the core and asymptotically, while the reconstructed lagrangian takes different limits. This Hamiltonian also has a non-Maxwell weak-field limit. For the electric branch $\mathcal P\to0^-$ one has $\mathcal H\propto(-\mathcal P)^{3/2}$, $\mathcal F=\mathcal P\mathcal H_{\mathcal P}^{2}\propto\mathcal P^{3}$, and $\mathcal L=2\mathcal P\mathcal H_{\mathcal P}-\mathcal H\propto(-\mathcal F)^{1/2}$ (up to convention-dependent normalization). The parametric electric invariant vanishes at both $r\to0$ and $r\to\infty$, so a globally single-valued Maxwell-like $\mathcal L(\mathcal F)$ is not obtained by this reconstruction. It is therefore a candidate \emph{effective} completion, not a globally regular microscopic gauge theory.

\paragraph{Scope of the background assumption.}
The noncommutative parameter $q=\sqrt\theta$ labels a prescribed Lorentzian-smeared family of classical metrics. Varying $q$ at fixed $M$ in our complexity plots is a comparison between geometries, not a thermodynamic variation proven to arise within one fixed-coupling matter theory. The metric satisfies Einstein's equations with the reconstructed anisotropic stress tensor, but this on-shell consistency is logically weaker than specifying an action, proving a well-posed variational principle, and determining all matter boundary counterterms. None of those additional assumptions enters the radial test-particle evolution or the two-scale obstruction.

\subsection{Energy, first-law identity, and its limitation}

At fixed asymptotic conventions, the constant term $-M$ in Eq.~\eqref{eq:asymptotic_holo} motivates the gravitational charge $E_{\rm grav}=M/(8G)$, up to a choice of AdS reference energy and possible matter boundary contributions. Put $h(r,q)=1-q/\sqrt{r^2+q^2}$. On the original geometric family, the horizon equation is $M=r_+^2/[l^2h(r_+,q)]$. With $T_H=f_\theta'(r_+)/4\pi$ and the Einstein-area entropy $S_{\rm BH}=\pi r_+/(2G)$ (natural units), direct differentiation gives
\begin{equation}
\begin{split}
 d\left(\frac{M}{8G}\right)&-T_HdS_{\rm BH}\\
 &=\frac{1}{8G}\left[\frac{q\,dM}{\sqrt{r_+^2+q^2}}
 +\frac{Mr_+^2\,dq}{(r_+^2+q^2)^{3/2}}\right].
\end{split}
 \label{eq:firstlaw_residual}
\end{equation}
The right-hand side is an exact \emph{algebraic residual identity}, not yet an independently derived thermodynamic work term. To see this directly, differentiating the horizon constraint $r_+^2/l^2-Mh(r_+,q)=0$ gives $f_\theta'(r_+)\,dr_+=h(r_+,q)\,dM+M\,\partial_q h(r_+,q)\,dq$; substitution into $T_HdS_{\rm BH}=f_\theta'(r_+)dr_+/(8G)$ reproduces Eq.~\eqref{eq:firstlaw_residual} without assuming a matter charge or a first law. Varying $M$ at fixed $q$ changes the amplitude of the effective matter distribution, so the vacuum BTZ first law need not apply to this restricted geometric ansatz.

Consider a formally enlarged metric family of the form $f(r)=r^2/l^2-M_0+C(Q)/\sqrt{r^2+s(Q)^2}$, with independent gravitational integration constant $M_0$ and electric integration constant $Q$. If this family were generated by a consistent fixed-coupling action, its horizon equation would imply a \emph{formal} differential identity $d[M_0/(8G)]=T_HdS_{\rm BH}+\Phi_QdQ$, where $\Phi_Q=(8G)^{-1}\partial_Q[C(Q)/\sqrt{r_+^2+s(Q)^2}]$ at fixed $r_+$ and fixed action couplings. A physical first law requires, in addition, a consistent gauge-charge normalization, potential and boundary terms derived from the same action. We have not constructed those ingredients or a microscopic CFT dual.

These observations refine the original question: a thermofield interpretation based solely on the shifted temperature would omit bulk matter data whose thermodynamic and boundary interpretation remains to be fixed. A generalized or excited TFD is a plausible candidate.

\section{Proper radial momentum and the momentum--Krylov prescription}\label{sec:formalism}

Following Refs.~\cite{Caputa:2026,Fan:2024iop,Fatemiabhari:2026}, consider a massive probe moving radially in a static geometry
\begin{equation}
 ds^2=-A(r)dt^2+A(r)B(r)dr^2+g_{ij}dv^i dv^j.
 \label{eq:generalmetric}
\end{equation}
With $t$ as worldline parameter, the particle action is
\begin{equation}
 S_{\rm p}=-m\int dt\,\sqrt{A(r)\left[1-B(r)\dot r^2\right]}.
\end{equation}
The conserved energy and radial velocity are
\begin{equation}
 E=\frac{m\sqrt{A(r)}}{\sqrt{1-B(r)\dot r^2}},
\end{equation}
\begin{equation}
 \dot r=\pm\frac{1}{\sqrt{B(r)}}
 \sqrt{1-\frac{m^2A(r)}{E^2}}.
 \label{eq:rdotgeneral}
\end{equation}
The momentum entering the geometric prescription is conjugate to proper radial distance $d\rho=\sqrt{A B}\,dr$. It is therefore
\begin{equation}
 p_\rho=\frac{m\sqrt{B(r)}\dot r}{\sqrt{1-B(r)\dot r^2}}.
 \label{eq:propermomentum}
\end{equation}
For a particle released from rest at $r_{\rm UV}$, $E=m\sqrt{A(r_{\rm UV})}$ and
\begin{equation}
 \dot r=-\frac{1}{\sqrt{B(r)}}
 \sqrt{1-\frac{A(r)}{A(r_{\rm UV})}},
 \label{eq:rdotUV}
\end{equation}
\begin{equation}
 \dot{\mathcal C}=|p_\rho|
 =m\sqrt{\frac{A(r_{\rm UV})}{A(r)}-1}.
 \label{eq:Cdotgeneral}
\end{equation}
The overall normalization of $\mathcal C$ depends on the boundary convention. Our focus is the geometric time dependence encoded in Eq.~\eqref{eq:Cdotgeneral}.

\section{Momentum--Krylov dynamics in the smeared BTZ black hole}\label{sec:application}

For Eq.~\eqref{eq:metric},
\begin{equation}
 A(r)=f_\theta(r),\qquad B(r)=\frac{1}{f_\theta(r)^2}.
\end{equation}
A probe released from rest at $r_{\rm UV}>r_+$ therefore obeys
\begin{equation}
 \dot r=-f_\theta(r)
 \sqrt{1-\frac{f_\theta(r)}{f_\theta(r_{\rm UV})}},
 \label{eq:rdotfinal}
\end{equation}
\begin{equation}
 \frac{|p_\rho|}{m}
 =\sqrt{\frac{f_\theta(r_{\rm UV})}{f_\theta(r)}-1}.
 \label{eq:Cdotfinal}
\end{equation}
The coordinate-time trajectory follows from
\begin{equation}
 t(r)=\int_r^{r_{\rm UV}}
 \frac{dr'}{f_\theta(r')\sqrt{1-f_\theta(r')/f_\theta(r_{\rm UV})}}.
 \label{eq:tr}
\end{equation}

Define the dimensionless time
\begin{equation}
 \tau=\frac{r_h}{l^2}t,
\end{equation}
and $X=x_{\rm UV}$. Equations~\eqref{eq:rdotfinal} and \eqref{eq:Cdotfinal} become
\begin{equation}
 \frac{dx}{d\tau}=-F_\alpha(x)
 \sqrt{1-\frac{F_\alpha(x)}{F_\alpha(X)}},
 \label{eq:dimensionlessODE}
\end{equation}
\begin{equation}
 \frac{|p_\rho|}{m}
 =\sqrt{\frac{F_\alpha(X)}{F_\alpha(x)}-1}.
 \label{eq:dimensionlessC}
\end{equation}
Thus, for fixed release point $X$, the shape of the evolution is governed by the single geometric deformation parameter $\alpha$; the probe mass only fixes the overall momentum scale.

\subsection{Exact BTZ solution}

\begin{figure}[htbp]
 \centering
  \includegraphics[width=\columnwidth]{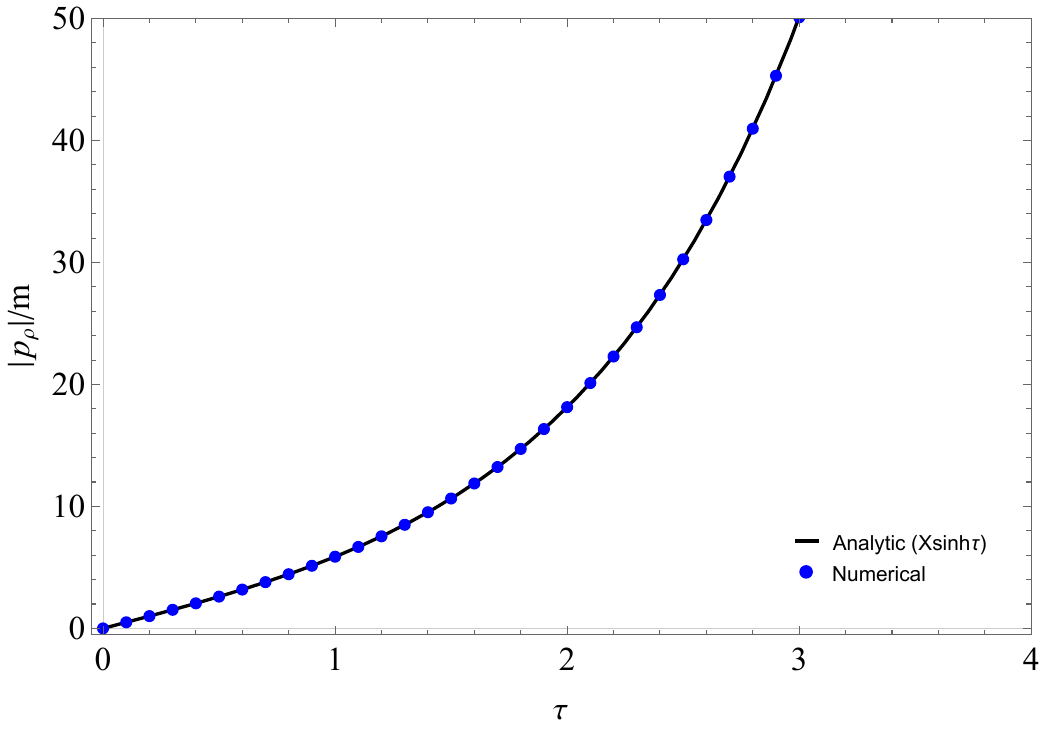}
  \caption{\footnotesize{Probe released from rest at $X=5$ in BTZ: numerical proper momentum and the exact result $|p_\rho|/m=5\sinh\tau$.}}
  \label{fig:btzexact}
 \end{figure}

The commutative BTZ limit provides a closed-form check. For $\alpha=0$, $F_0=x^2-1$. A particle released from rest at $X$ satisfies
\begin{equation}
 \frac{|p_\rho|}{m}=X\sinh\tau,
 \label{eq:BTZPexact}
\end{equation}
\begin{equation}
 x^2(\tau)=1+\frac{X^2-1}{1+X^2\sinh^2\tau}.
 \label{eq:BTZxexact}
\end{equation}
For the numerical choice $X=5$, $|p_\rho|/m=5\sinh\tau$. Figure~\ref{fig:btzexact} shows the numerical solution superposed on the exact expression.

\subsection{Lorentzian smearing and radial infall}

\begin{figure*}[t]
 \centering
  \subfloat[]{\includegraphics[width=0.45\textwidth]{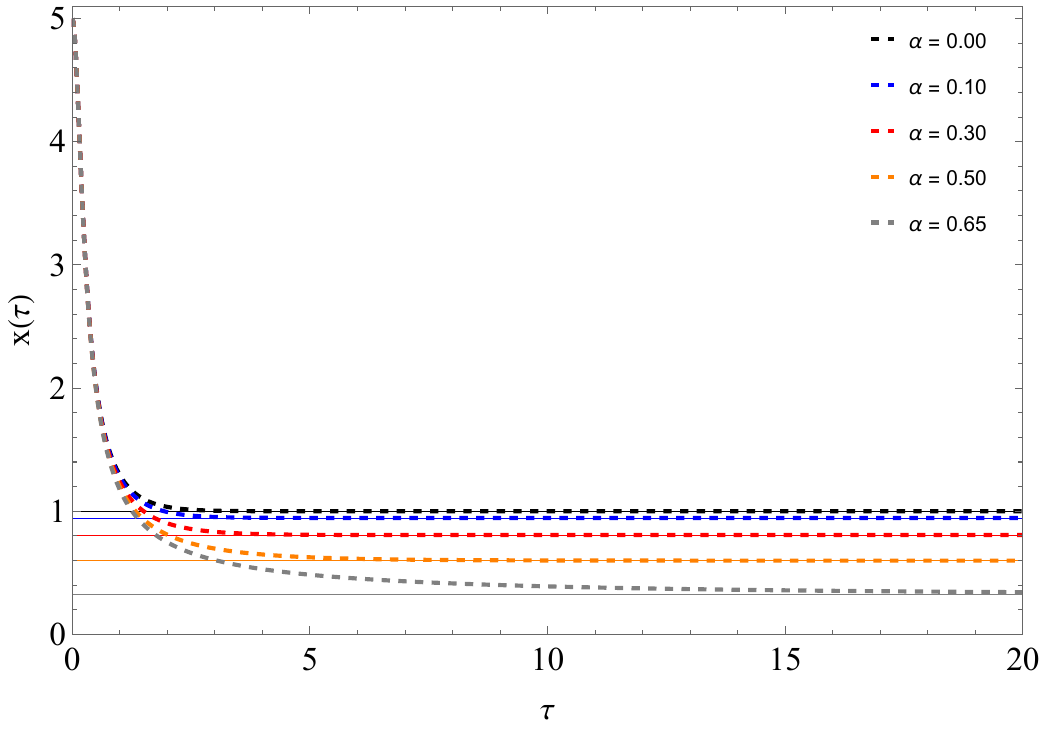}\label{fig:trajectory}}
  \quad
   \subfloat[]{\includegraphics[width=0.45\textwidth]{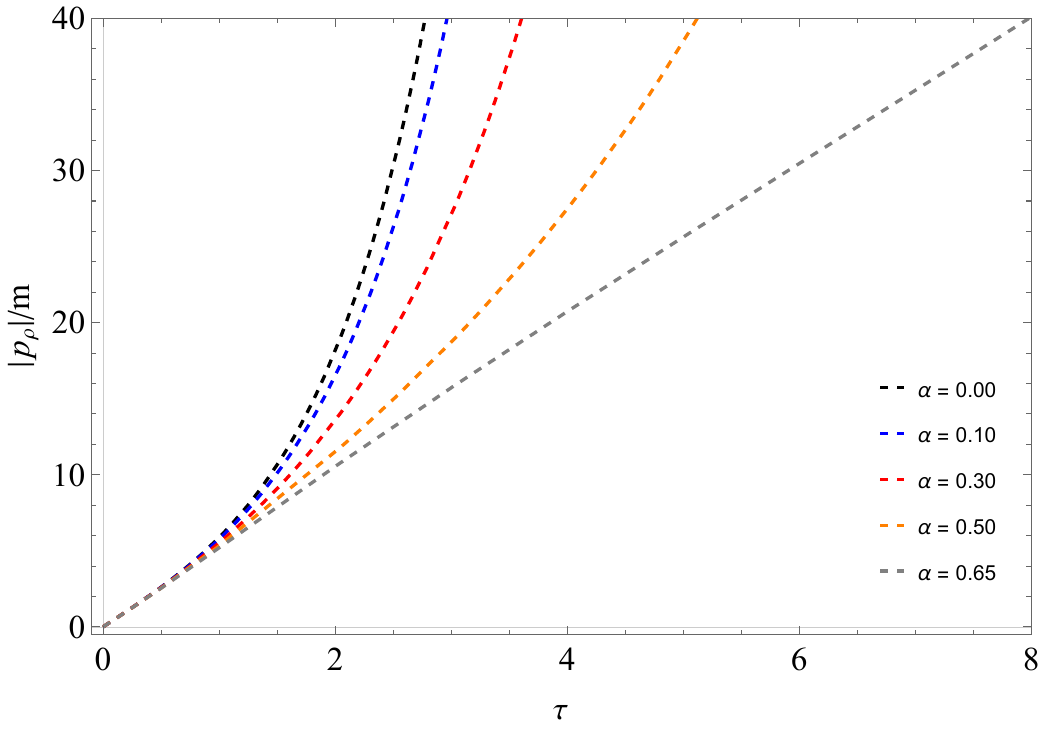}\label{fig:complexityrate}}
  \caption{\footnotesize{In (a) radial infall from rest at $X=5$ for representative Lorentzian smearings. Increasing $\alpha$ shifts the finite-radius outer horizon inward and delays its approach in the dimensionless coordinate time $\tau$. In (b) proper radial momentum $|p_\rho|/m$, interpreted through the momentum-Krylov prescription as the geometric complexity-growth rate up to an overall normalization. Lorentzian smearing suppresses the late-time growth.}}
 \end{figure*}

We integrate Eq.~\eqref{eq:dimensionlessODE} for
\begin{equation}
 X=5,\qquad \alpha=0,\;0.1,\;0.3,\;0.5,\;0.65,
\end{equation}
all within the finite-radius black-hole domain $\alpha<1/\sqrt2$. The trajectory plots and the late-time exponent extraction use different numerical ranges. The figures display the finite-time radial evolution, whereas the exponent is fitted separately from a near-horizon integration as specified below. The plotted trajectories are integrated until $x=x_++\delta$; a check at $\alpha=0.3$ gives overlapping curves for $\delta=10^{-2},10^{-3},10^{-4}$ away from the terminal region, so the displayed finite-radius evolution is stable under this stopping prescription. No such cutoff is imposed on the exponent extraction. Figure~\ref{fig:trajectory} shows that increasing $\alpha$ moves the outer horizon inward (horizontal lines), while the corresponding fall trajectory asymptotically approaches its horizon at increasingly later values of the dimensionless BTZ coordinate time $\tau$.

The corresponding proper momentum is shown in Fig.~\ref{fig:complexityrate}. All curves start at zero because the probe is released from rest. The deformation becomes increasingly visible at late times, where the growth rate is reduced as $\alpha$ increases.
\subsection{Surface gravity and late-time growth}

Near a nondegenerate outer horizon, writing $x=x_++\Delta x$ gives
\begin{equation}
 F_\alpha(x)=F_\alpha'(x_+)\Delta x+O(\Delta x^2).
\end{equation}
Equation~\eqref{eq:dimensionlessODE} then implies
\begin{equation}
 \Delta x\propto e^{-F_\alpha'(x_+)\tau},
\end{equation}
and Eq.~\eqref{eq:dimensionlessC} yields
\begin{equation}
 \frac{|p_\rho|}{m}\propto e^{\hat\kappa\tau},
 \qquad
 \hat\kappa=\frac12F_\alpha'(x_+),
 \label{eq:lateK}
\end{equation}
where
\begin{equation}
 F_\alpha'(x)=2x-
 \frac{\alpha x}{(x^2+\alpha^2)^{3/2}}.
 \label{eq:Fprime}
\end{equation}
The physical surface gravity is $\kappa=(r_h/l^2)\hat\kappa$. The Hawking scale of the smeared BTZ geometry thus sets the late-time exponent of the proper-momentum prescription. This relation is the expected near-horizon behavior of the momentum construction \cite{Fan:2024iop}; the model-dependent information is the deformation of $x_+$ and $\kappa$ by $\alpha$.

\begin{figure*}[t]
 \centering
  \subfloat[]{\includegraphics[width=0.45\textwidth]{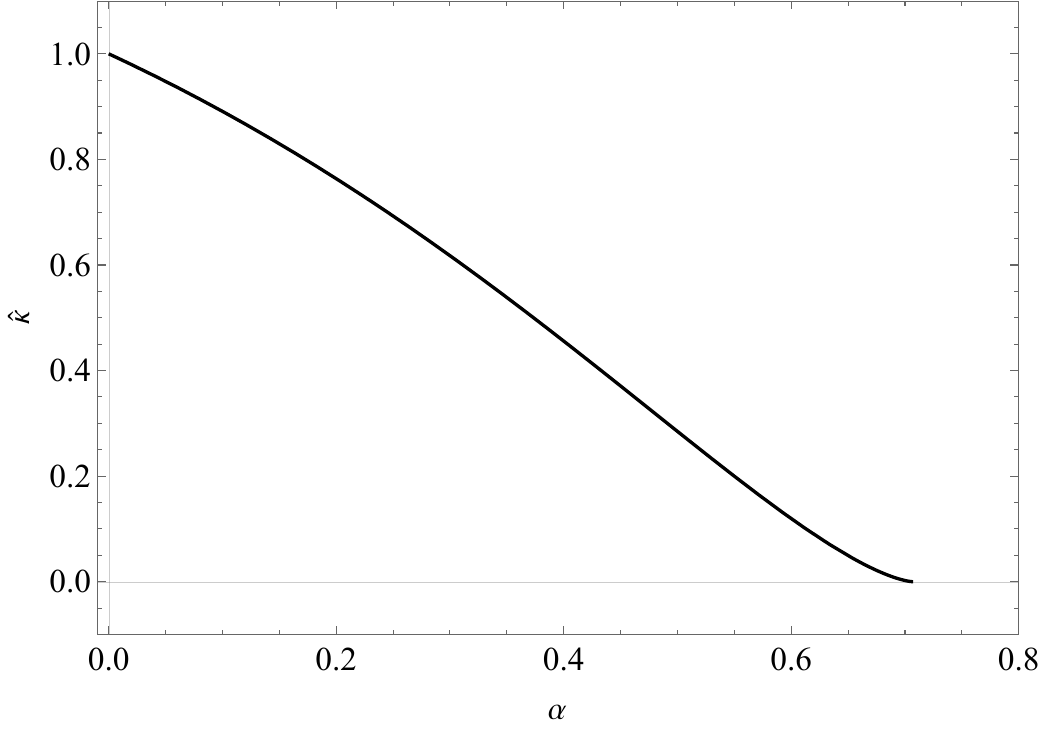}\label{fig:kappa}}
  \quad
   \subfloat[]{\includegraphics[width=0.45\textwidth]{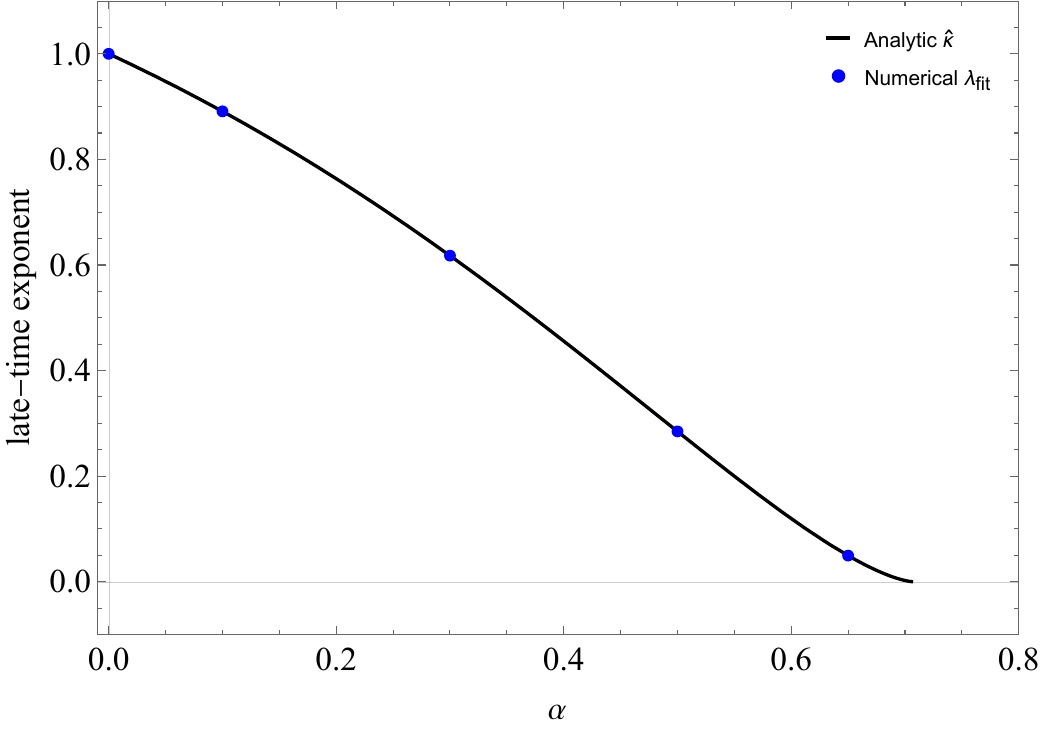}\label{fig:kappafit}}
  \caption{\footnotesize{In (a) dimensionless surface gravity $\hat\kappa=F_\alpha'(x_+)/2$ of the finite-radius outer horizon. Lorentzian smearing monotonically lowers the surface-gravity scale controlling the late-time proper-momentum growth. In (b) analytic near-horizon exponent $\hat\kappa$ and numerical fit $\lambda_{\rm fit}$. Their agreement verifies the surface-gravity control of the late-time proper-momentum growth.}}
 \end{figure*}

Figure~\ref{fig:kappa} shows the monotonic suppression of $\hat\kappa$. It starts at the BTZ value $\hat\kappa=1$ and tends to zero as the finite-radius horizon approaches the critical endpoint.

To test Eq.~\eqref{eq:lateK} quantitatively, we fit
\begin{equation}
 \log(|p_\rho|/m)=\lambda_{\rm fit}\tau+c
\end{equation}
in the near-horizon shell $10^{-7}<x-x_+<10^{-4}$. To obtain the fit independently of the plotting interval, we use the exact first integral $dx/d\tau=-F_\alpha(x)\sqrt{1-F_\alpha(x)/F_\alpha(X)}$ with $X=5$. The coordinate time is evaluated by quadrature,
\begin{equation}
\tau(x)=\int_x^X\!\frac{dy}{F_\alpha(y)\sqrt{1-F_\alpha(y)/F_\alpha(X)}},
\end{equation}
with the integrable square-root singularity at $y=X$ removed by the substitution $y=X-u^2$. We sample 35 logarithmically spaced values of $x-x_+$ between $10^{-7}$ and $10^{-4}$, compute $|p_\rho|/m=\sqrt{F_\alpha(X)/F_\alpha(x)-1}$, and perform an unweighted linear least-squares fit of $\log(|p_\rho|/m)$ against $\tau$. This reproduces the tabulated exponents to the quoted precision up to small sampling-dependent differences (of order $10^{-6}$). The corresponding fitting-time ranges are approximately $[4.93,8.39]$, $[5.45,9.32]$, $[7.47,13.06]$, $[14.71,26.84]$, and $[71.84,141.69]$ for $\alpha=0,0.1,0.3,0.5,0.65$, respectively. The $\alpha=0.65$ fit extends well beyond the time range displayed in the trajectory figure. The results are shown in Table~\ref{tab:kappafit}.
\begin{table}[t]
\caption{Analytic surface-gravity exponent and numerical late-time fit.}
\label{tab:kappafit}
\begin{ruledtabular}
\begin{tabular}{cccc}
$\alpha$ & $x_+$ & $\hat\kappa$ & $\lambda_{\rm fit}$ \\
\hline
0.00 & 1.000000 & 1.000000 & 1.000009 \\
0.10 & 0.945978 & 0.891028 & 0.891038 \\
0.30 & 0.807245 & 0.617660 & 0.617672 \\
0.50 & 0.599676 & 0.284701 & 0.284713 \\
0.65 & 0.324474 & 0.049444 & 0.049450 \\
\end{tabular}
\end{ruledtabular}
\end{table}
The tabulated relative discrepancy is of order $10^{-2}\%$ or smaller for the sampled deformations; the independent verification gives approximately $1.26\times10^{-2}\%$ at $\alpha=0.65$ because of the sampling and regression choices. Figure~\ref{fig:kappafit} displays the analytic and fitted exponents together.
\subsection{Momentum growth and the MSS thermal scale}\label{sec:mss}

The surface-gravity result also permits a direct comparison with the thermal scale that appears in the MSS bound on quantum chaos. Restoring $k_B$ and $\hbar$, the Hawking temperature of the finite-radius outer horizon is
\begin{equation}
 T_H(\alpha)=\frac{\hbar\kappa(\alpha)}{2\pi k_B},
 \label{eq:hawkingMSS}
\end{equation}
where $\kappa=(r_h/l^2)\hat\kappa$. On the other hand, the late-time logarithmic growth rate of the geometric momentum observable is
\begin{equation}
 \lambda_{\rm geom}\equiv
 \lim_{t\to\infty}\frac{d}{dt}\log\frac{|p_\rho|}{m}
 =\kappa(\alpha).
 \label{eq:lambdageom}
\end{equation}
Combining Eqs.~\eqref{eq:hawkingMSS} and \eqref{eq:lambdageom} gives
\begin{equation}
 \boxed{\lambda_{\rm geom}(\alpha)=\kappa(\alpha)
 =\frac{2\pi k_B T_H(\alpha)}{\hbar}}.
 \label{eq:mssscale}
\end{equation}
The right-hand side is the maximal thermal scale entering the MSS bound
\begin{equation}
 \lambda_L\leq\lambda_{\rm MSS}^{\max}
 \equiv\frac{2\pi k_B T}{\hbar}
 \label{eq:MSSbound}
\end{equation}
for the Lyapunov exponent extracted from a thermal out-of-time-order correlator \cite{Maldacena:2015waa}. The momentum--Krylov probe exponent therefore lies at the MSS thermal scale when the temperature is identified with the Hawking temperature of the same geometry,
\begin{equation}
 \lambda_{\rm geom}=\lambda_{\rm MSS}^{\max}(T_H).
 \label{eq:scaleequality}
\end{equation}
Equation~\eqref{eq:scaleequality} is an equality of geometric and thermal scales. It is \emph{not} a derivation of $\lambda_L=\lambda_{\rm geom}$ for the boundary theory: showing saturation of the MSS chaos bound would require an OTOC calculation (or an equivalent microscopic chaos indicator) for the theory dual to the Lorentzian-smeared background. This qualification is consistent with the broader Krylov literature, where exponential Krylov growth and conventional Lyapunov indicators need not be interchangeable \cite{Avdoshkin:2022xuw,Hashimoto:2023swv}.

\begin{figure}[htbp]
 \centering
  \includegraphics[width=\columnwidth]{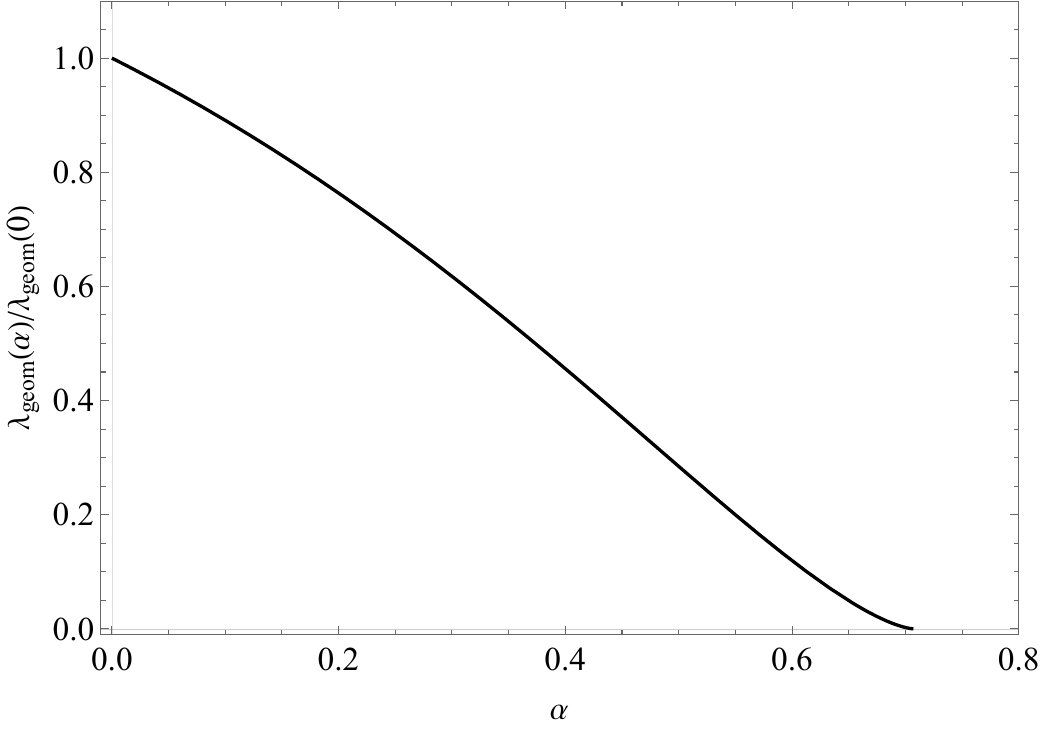}
  \caption{\footnotesize{Normalized geometric momentum exponent. Since $\lambda_{\rm geom}=\kappa=2\pi k_BT_H/\hbar$, the same curve also gives $T_H(\alpha)/T_H(0)$ and the normalized MSS thermal scale.}}
  \label{fig:mssscale}
 \end{figure}

In normalized form,
\begin{equation}
 \frac{\lambda_{\rm geom}(\alpha)}{\lambda_{\rm geom}(0)}
 =\frac{T_H(\alpha)}{T_H(0)}
 =\frac{\kappa(\alpha)}{\kappa(0)}
 =\hat\kappa(\alpha),
 \label{eq:normalizedMSS}
\end{equation}
where $\hat\kappa(0)=1$. Figure~\ref{fig:mssscale} displays simultaneously the suppression of the geometric momentum exponent, Hawking temperature and corresponding MSS thermal scale by Lorentzian smearing.

As $\alpha\to\alpha_c^-=1/\sqrt2$, the finite-radius horizon approaches the degenerate core and $\kappa$, $T_H$, and $\lambda_{\rm geom}$ all tend to zero. Exactly at the critical endpoint the usual nonextremal argument leading to Eq.~\eqref{eq:lateK} ceases to apply: the horizon is no longer a simple zero of $F_\alpha$. The vanishing limit is approached from the nondegenerate black-hole branch and says nothing about exponential growth at the critical geometry itself.

\section{Finite-time noncommutative response of momentum--Krylov growth}\label{sec:thetaresponse}

The proper-momentum prescription gives a direct way to measure the effect of the Lorentzian smearing parameter. With the probe mass and the overall momentum normalization held fixed, define the dimensionless accumulated geometric observable
\begin{equation}
\begin{gathered}
 \mathcal K_\alpha(\tau)=\int_0^\tau d\tau'\,\frac{|p_\rho(\tau';\alpha)|}{m},\\
 \Delta\mathcal K_\alpha(\tau)=\mathcal K_\alpha(\tau)-\mathcal K_0(\tau).
\end{gathered}
 \label{eq:Kaccum}
\end{equation}
The integral is a geometric realization of the momentum--Krylov prescription; no microscopic Lanczos recursion is inferred from it. The same release point $X$, asymptotic mass $M$, AdS scale $l$, and dimensionless time $\tau$ are used for every $\alpha$. The BTZ result (\ref{eq:BTZPexact}) integrates exactly to
\begin{equation}
 \mathcal K_0(\tau)=X(\cosh\tau-1).
 \label{eq:KBTZexact}
\end{equation}

The early-time behavior can be obtained without numerical integration. Expanding the radial equation about release from rest gives $x(\tau)=X-F_\alpha(X)F'_\alpha(X)\tau^2/4+O(\tau^4)$, and consequently
\begin{equation}
\begin{gathered}
 \frac{|p_\rho|}{m}=\frac{F'_\alpha(X)}2\tau+O(\tau^3),\\
 \mathcal K_\alpha(\tau)=c_2(\alpha)\tau^2+O(\tau^4),
 \quad c_2(\alpha)=\frac{F'_\alpha(X)}4.
\end{gathered}
 \label{eq:Kearly}
\end{equation}
For the Lorentzian profile,
\begin{equation}
 F'_\alpha(X)=2X-\frac{\alpha X}{(X^2+\alpha^2)^{3/2}}.
\end{equation}
The early-time coefficient probes the local metric derivative at the release point, whereas the late-time logarithmic slope of the momentum probes $F'_\alpha(x_+)/2$. The two quantities are not proportional as $\alpha$ varies.

\begin{figure}[htbp]
\centering
\includegraphics[width=\columnwidth]{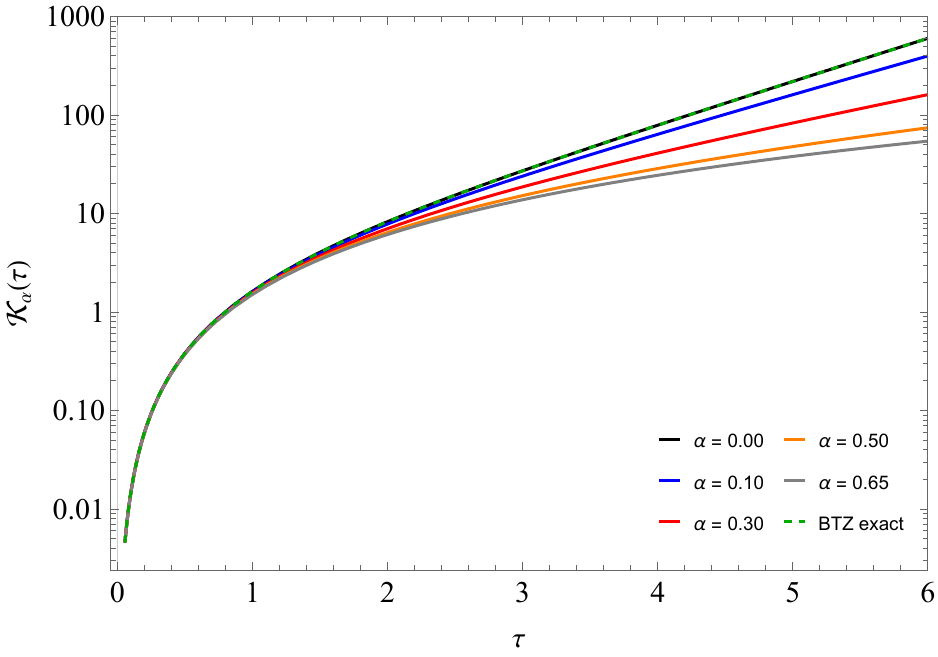}
\caption{\footnotesize{Accumulated geometric momentum--Krylov observable $\mathcal K_\alpha(\tau)$ at fixed $M=l=m=1$, release point $X=3$ and common normalization, for $\alpha=0,0.1,0.3,0.5,0.65$. The dashed reference is the analytic BTZ expression $3(\cosh\tau-1)$.}}
\label{fig:thetaK}
\end{figure}

The trajectory and near-horizon-fit indicators in Sec.~\ref{sec:application} use $X=5$; the accumulated-response and two-scale comparisons below instead use $X=3$. For the finite-time comparison we choose $M=l=m=1$ and $X=3$ and integrate Eqs.~(\ref{eq:dimensionlessODE}) and (\ref{eq:Kaccum}) with $\mathcal K_\alpha(0)=0$. The finite-time data are displayed in Fig.~\ref{fig:thetaK}. The parameter is $\alpha=\sqrt\theta/(l\sqrt M)$, so varying $\alpha$ at fixed $M,l$ varies $\theta$ without changing the comparison ensemble.

To distinguish finite-time deformation from the trivial replacement of the late-time exponent, consider the parameter-free, time-only rescaling hypothesis
\begin{equation}
 \mathcal K_\alpha(\tau)\stackrel{?}{=}\mathcal K_0\!\left(\frac{\hat\kappa_\alpha}{\hat\kappa_0}\tau\right).
 \label{eq:rescalehyp}
\end{equation}
It fails already at short times: Eq.~(\ref{eq:Kearly}) would require $c_2(\alpha)=c_2(0)(\hat\kappa_\alpha/\hat\kappa_0)^2$, whereas $c_2(\alpha)$ changes only weakly for $X=3$ even as $\hat\kappa_\alpha$ approaches zero. This rejects the \emph{specific} time-only collapse (\ref{eq:rescalehyp}); it does not exclude a more general parameter-dependent amplitude and time redefinition. The key result is a quantitatively distinct early- and late-time response to $\theta$ within the same geometric prescription.

\subsection{An analytic obstruction to amplitude--time collapse}\label{subsec:two_scale}
The previous test fixes the amplitude to unity. We now allow the more general two-parameter hypothesis
\begin{equation}
 \mathcal K_\alpha(\tau)\stackrel{?}{=}A_\alpha\mathcal K_0(B_\alpha\tau),\qquad A_\alpha>0,\ B_\alpha>0,
 \label{eq:ABcollapse}
\end{equation}
with constants independent of time. This is more general than the time-only rescaling in Eq.~\eqref{eq:rescalehyp}, but it does not cover arbitrary nonlinear time reparametrizations.

Writing $F_j=\partial_x^j F_\alpha(X)$, the release-from-rest equations yield the analytic expansion
\begin{equation}
\begin{gathered}
 \mathcal K_\alpha(\tau)=c_2(\alpha)\tau^2+c_4(\alpha)\tau^4+O(\tau^6),\\
 c_2=\frac{F_1}{4},\qquad
 \boxed{c_4=\frac{F_1^3-2F_0F_1F_2}{192}}.
\end{gathered}
 \label{eq:Kquartic}
\end{equation}
To see this, put $\delta=X-x=a\tau^2+b\tau^4+O(\tau^6)$ in the radial equation. One obtains
\begin{equation}
 a=\frac{F_0F_1}{4},\qquad
 b=-\frac{a}{12}\left(2F_1^2+\frac{F_0F_2}{2}\right),
\end{equation}
which, when substituted into $|p_\rho|/m=\sqrt{F_0/F_\alpha(x)-1}$ and integrated, gives Eq.~\eqref{eq:Kquartic}. For the undeformed BTZ background at $X=3$, $c_2(0)=3/2$ and $c_4(0)=1/8$, in agreement with $3(\cosh\tau-1)$.

The perturbative mismatch follows without fitting numerical trajectories. At fixed $X>1$ and for $\alpha\to0^+$, the Lorentzian profile has the nonanalytic-in-$\theta$ expansion
\begin{equation}
\begin{split}
 F_\alpha(X)&=X^2-1+\frac{\alpha}{X}+O(\alpha^3),\\
 F_1&=2X-\frac{\alpha}{X^2}+O(\alpha^3),\\
 F_2&=2+\frac{2\alpha}{X^3}+O(\alpha^3).
\end{split}
 \label{eq:smallalphaF}
\end{equation}
Consequently, direct substitution into Eq.~\eqref{eq:Kquartic} gives
\begin{equation}
 \frac{c_4(\alpha)}{c_2(\alpha)}=\frac1{12}-\frac{3X^2-1}{12X^3}\alpha+O(\alpha^2).
 \label{eq:smallalphaRatio}
\end{equation}
The positive outer horizon is $x_+=1-\alpha/2+O(\alpha^2)$, and its surface gravity is $\hat\kappa_\alpha=1-\alpha+O(\alpha^2)$. The early-time and late-time scales therefore disagree already at first order:
\begin{equation}
 \boxed{\begin{aligned}
 B_{\rm early}&=1-\frac{3X^2-1}{2X^3}\alpha+O(\alpha^2),\\
 B_{\rm late}&=1-\alpha+O(\alpha^2).
 \end{aligned}}
 \label{eq:perturbativeMismatch}
\end{equation}
For the numerical release radius $X=3$, $B_{\rm early}-B_{\rm late}=\frac{14}{27}\alpha+O(\alpha^2)>0$ for sufficiently small positive $\alpha$. More generally the leading difference is $[1-(3X^2-1)/(2X^3)]\alpha$, which is strictly positive for $X>1$ since $2X^3-3X^2+1=(X-1)^2(2X+1)>0$. This shows the failure of an exact two-parameter collapse perturbatively for any fixed exterior release point $X>1$ and sufficiently small nonzero deformation. It does not assert a uniform lower bound for arbitrary $\alpha$ or arbitrary initial conditions.

Matching the early $\tau^2$ and $\tau^4$ coefficients in Eq.~\eqref{eq:ABcollapse} uniquely fixes
\begin{equation}
\begin{gathered}
 B_{\rm early}^2(\alpha)=\frac{c_4(\alpha)/c_2(\alpha)}{c_4(0)/c_2(0)}
 =12\frac{c_4(\alpha)}{c_2(\alpha)},\\
 A_{\rm early}=\frac{c_2(\alpha)}{c_2(0)B_{\rm early}^2},
\end{gathered}
 \label{eq:Bearly}
\end{equation}
provided $c_4(\alpha)>0$, as holds for the sampled range. Separately, matching the late-time exponential growth of $\mathcal K$ requires
\begin{equation}
 B_{\rm late}(\alpha)=\frac{\hat\kappa_\alpha}{\hat\kappa_0}=\hat\kappa_\alpha.
 \label{eq:Blate}
\end{equation}
The amplitude $A_\alpha$ cannot alter this exponent. Hence $B_{\rm early}\ne B_{\rm late}$ is an analytic obstruction to \emph{any} exact constant-amplitude, constant-time rescaling of the BTZ accumulated-momentum curve over both regimes. This conclusion does not require fitting the full numerical trajectory.

\begin{table}[htbp]
\centering
\caption{Analytic test of two-parameter BTZ curve collapse for $X=3$, $M=l=m=1$. $B_{\rm early}$ follows from the quadratic and quartic coefficients; $B_{\rm late}$ follows independently from the near-horizon exponent. Their disagreement excludes Eq.~\eqref{eq:ABcollapse}.}
\label{tab:ABtest}
\begin{tabular}{ccccc}
\hline
$\alpha$ & $c_4(\alpha)$ & $B_{\rm early}$ & $B_{\rm late}$ & $B_{\rm early}/B_{\rm late}$ \\
\hline
$0$ & $0.125000$ & $1.000000$ & $1.000000$ & $1.000$ \\
$0.10$ & $0.112764$ & $0.950676$ & $0.891028$ & $1.067$ \\
$0.30$ & $0.088874$ & $0.845519$ & $0.617660$ & $1.369$ \\
$0.50$ & $0.066545$ & $0.732894$ & $0.284701$ & $2.574$ \\
$0.65$ & $0.051291$ & $0.644199$ & $0.049444$ & $13.029$ \\
\hline
\end{tabular}
\end{table}

\paragraph{Size of the shape distortion.}
The ratio $B_{\rm early}/B_{\rm late}$ compares characteristic time scales and is not itself the fractional distortion of $\mathcal K$. With $B=\hat\kappa_\alpha$ and $A=c_2(\alpha)/[c_2(0)B^2]$, the normalized shape ratio $\mathcal K_\alpha/[A\mathcal K_0(B\tau)]$ approaches approximately $1.017$, $1.068$, $1.108$ and $0.872$ at late times for $\alpha=0.1,0.3,0.5,0.65$, respectively, for $X=3$. The finite-time shape effect is modest, although the analytic obstruction is exact. For near-critical parameters, times of order $1/\hat\kappa_\alpha$ are needed to resolve the late-time exponent clearly; the plotted interval $\tau\leq14$ is not such an asymptotic window for $\alpha=0.65$.

\begin{figure}[htbp]
\centering
\includegraphics[width=\columnwidth]{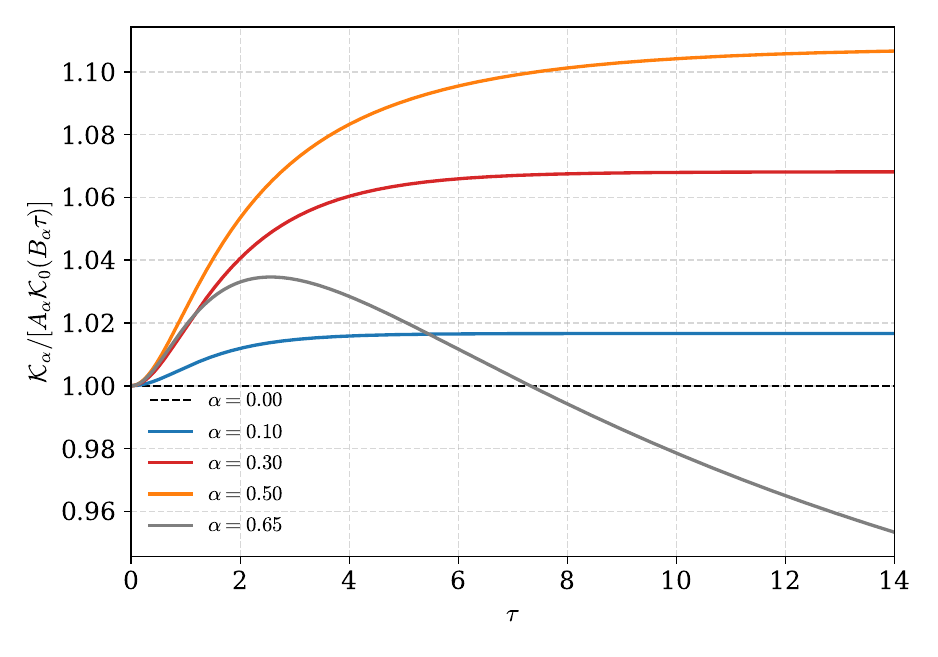}
\caption{\footnotesize{Finite-time test of the amplitude--time collapse with the late-time exponent $B_\alpha=\hat\kappa_\alpha$ and the amplitude $A_\alpha$ fixed by the exact early quadratic coefficient. All ratios tend to one at the origin; their departures measure changes of curve shape that these two scales cannot remove. The analytic mismatch of $B_{\rm early}$ and $B_{\rm late}$ rules out any exact two-parameter collapse.}}
\label{fig:ABtest}
\end{figure}

For a complementary finite-time indicator, impose the \emph{late} exponent and \emph{early} quadratic amplitude simultaneously: $B_\alpha=B_{\rm late}$ and $A_\alpha=c_2(\alpha)/[c_2(0)B_{\rm late}^2]$. The ratio $\mathcal K_\alpha(\tau)/[A_\alpha\mathcal K_0(B_\alpha\tau)]$ then tends to unity as $\tau\to0$, but its quartic deviation is proportional to $B_{\rm early}^2-B_{\rm late}^2$ and is nonzero for $\alpha>0$ in Table~\ref{tab:ABtest}. Figure~\ref{fig:ABtest} shows this mismatch without an arbitrary fitted amplitude.

The obstruction is a statement about the \emph{geometric} accumulated proper momentum for this fixed-release-point family.

The background is a noncommutative-\emph{inspired} Lorentzian mass distribution, not a derived noncommutative boundary field theory, so the calculation does not determine the microscopic Krylov coefficients of a noncommutative CFT.

\subsection{Asymptotic release and geometric UV/IR separation}\label{subsec:uv_ir}
At fixed nonzero smearing, the early-time scale approaches its BTZ value as the release point is moved towards the asymptotic region. Expanding Eq.~\eqref{eq:Bearly} at large $X$ gives
\begin{equation}
 B_{\rm early}=1-\frac{3\alpha}{2X}+O(X^{-2}),\qquad
 B_{\rm late}=\hat\kappa_\alpha<1.
 \label{eq:UVIR}
\end{equation}
More generally, writing $F(x)=x^2-1+\delta(x)$, the exact early-time coefficient implies
\begin{equation}
\begin{split}
 B_{\rm early}^2={}&1+X\delta'(X)-\delta(X)-\frac{X^2-1}{2}\delta''(X)\\
 &+\frac{\delta'(X)^2-2\delta(X)\delta''(X)}4.
\end{split}
 \label{eq:general_asymptotic_B}
\end{equation}
If $\delta(x)\sim cx^{-n}$ with $n>0$ and the first two derivatives admit the corresponding differentiated asymptotic expansions, then
\begin{equation}
 B_{\rm early}^2-1\sim-\frac{(n+1)(n+2)}2cX^{-n}.
\end{equation}
Hence $B_{\rm early}\to1$ as $X\to\infty$, while $B_{\rm late}=\hat\kappa_\alpha$ is independent of $X$. The mismatch tends to $1-\hat\kappa_\alpha=1-T_H/T_{\rm BTZ}(M)$, and vanishes in this limit only if the two Hawking temperatures coincide.

\begin{figure}[htbp]
\centering
\includegraphics[width=\columnwidth]{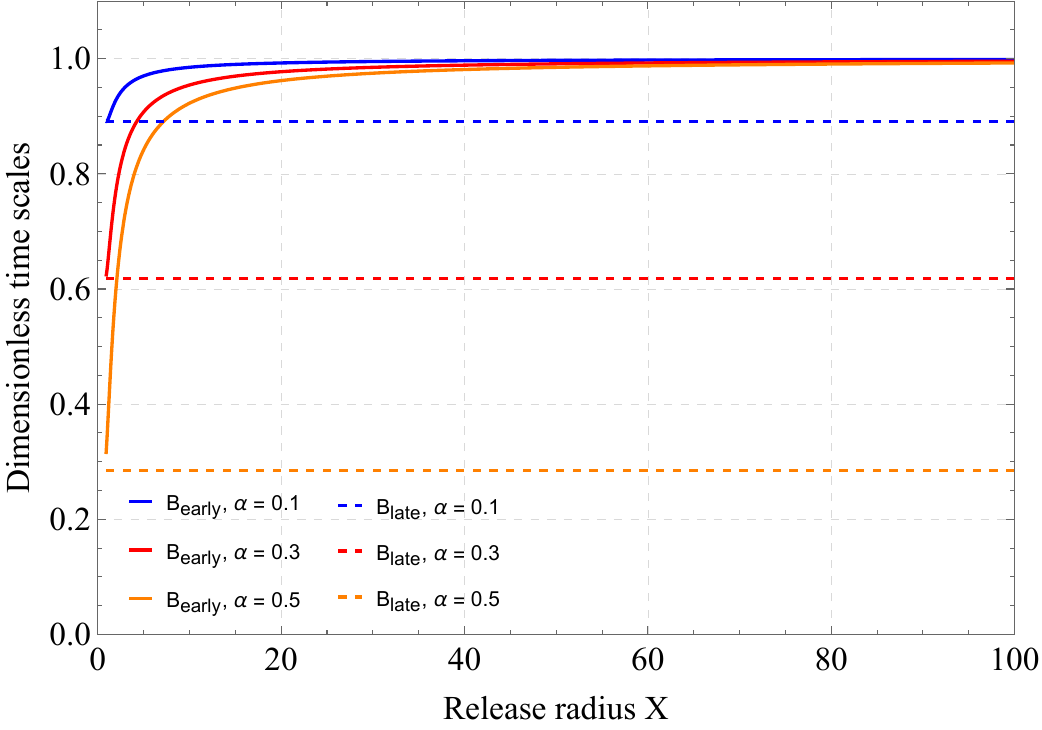}
\caption{\footnotesize{The early-time scale $B_{\rm early}$ (solid) approaches unity as the release radius $X$ increases, while the horizon-controlled scale $B_{\rm late}$ (dashed) remains fixed by $\alpha$. Solid and dashed curves with the same color correspond to the same $\alpha$; the dashed lines show $B_{\rm late}=\hat\kappa_\alpha$. All values follow analytically without fitting full trajectories.}}
\label{fig:uv_ir}
\end{figure}

The large-$X$ expansion is taken at fixed smearing and is distinct from the small-$\alpha$ expansion in Eq.~\eqref{eq:perturbativeMismatch}. For example, at $\alpha=0.3$, direct evaluation at $X=100$ yields $B_{\rm early}=0.99549$ while $B_{\rm late}=0.61766$. The initial response of a far-away probe thus barely resolves the noncommutative-inspired deformation, whereas the near-horizon growth rate retains an order-one sensitivity. Figure~\ref{fig:uv_ir} illustrates this behavior and makes contact with the unchanged gravitational $g_{(2)}$ coefficient and deformation-sensitive subleading data discussed in Sec.~\ref{sec:fg}.

\subsection{Comparison with BTZ at fixed mass and at fixed temperature}\label{subsec:btzcomparison}
The absence of an exact amplitude--time collapse is a useful analytic statement, but the physical interpretation is clearer with two fixed prescriptions. At the same asymptotic mass $M$ and initial dimensionless radius $X$, the undeformed reference is $\mathcal K_{\rm BTZ}^{(M)}=X(\cosh\tau-1)$. A second reference is the ordinary BTZ black hole with the \emph{same Hawking temperature} as the smeared background. In units where the original $M=l=1$, its BTZ mass is $M_{\rm ref}=\hat\kappa_\alpha^2$ and, for the same physical release radius $r_0$, the reference dimensionless release point is $X_{\rm ref}=X/\hat\kappa_\alpha$. Since its proper momentum is $|p_\rho|/m=X_{\rm ref}\sinh(\hat\kappa_\alpha\tau)$ in the common time $\tau$, integration gives
\begin{equation}
\begin{gathered}
 \mathcal K_{\rm BTZ}^{(T)}(\tau)=\frac{X}{\hat\kappa_\alpha^2}
 \left[\cosh(\hat\kappa_\alpha\tau)-1\right],\\
 R_M=\frac{\mathcal K_\alpha}{\mathcal K_{\rm BTZ}^{(M)}},\qquad
 R_T=\frac{\mathcal K_\alpha}{\mathcal K_{\rm BTZ}^{(T)}}.
\end{gathered}
 \label{eq:two_references}
\end{equation}
These reference curves have no fitted amplitude or time scale: their factors follow from the BTZ mass--temperature relation and the choice of the same physical $r_0$ and probe mass. For $X=3$, $\alpha=0.3$, we obtain $R_M=(0.988,0.850,0.268)$ and $R_T=(1.001,1.035,1.060)$ at $\tau=(0.5,2,6)$, respectively. At $\alpha=0.5$ the corresponding ratios are $R_M=(0.980,0.777,0.123)$ and $R_T=(0.999,1.044,1.086)$. At fixed mass the smeared response falls far below BTZ at late times, whereas the BTZ black hole with the same Hawking temperature reproduces it to within about $10\%$ over the displayed window ($\tau\le14$) for all sampled $\alpha$.

The two reference curves are related to the collapse test of Sec.~\ref{subsec:two_scale} in a simple way. Since $\mathcal K_{\rm BTZ}^{(T)}(\tau)=\hat\kappa_\alpha^{-2}\mathcal K_0(\hat\kappa_\alpha\tau)$, the matched-temperature reference is exactly of the form~\eqref{eq:ABcollapse} with $B_\alpha=B_{\rm late}$ and $A_\alpha=\hat\kappa_\alpha^{-2}$. Its early-time coefficient is the BTZ value $c_2(0)=X/2$, which differs from $c_2(\alpha)$ only by the factor $c_2(\alpha)/c_2(0)=0.995$ and $0.991$ for $\alpha=0.3$ and $0.5$. Hence $R_T$ coincides, up to this constant, with the shape ratio of Fig.~\ref{fig:ABtest}: the matched-temperature BTZ black hole is the physical realization of the late-exponent collapse, and the residual few-percent drift of $R_T$ (towards $1.062$ and $1.098$ at late times for $\alpha=0.3$ and $0.5$) is the finite-time imprint of the analytic obstruction $B_{\rm early}\neq B_{\rm late}$.
\begin{figure*}[t]
 \centering
  \subfloat[]{\includegraphics[width=0.45\textwidth]{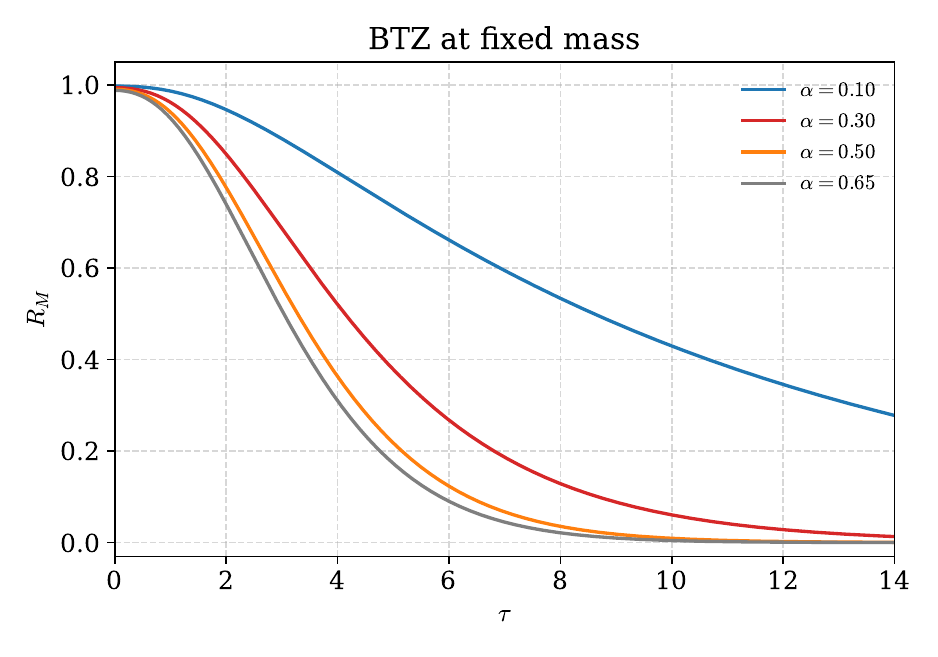}}
  \quad
   \subfloat[]{\includegraphics[width=0.45\textwidth]{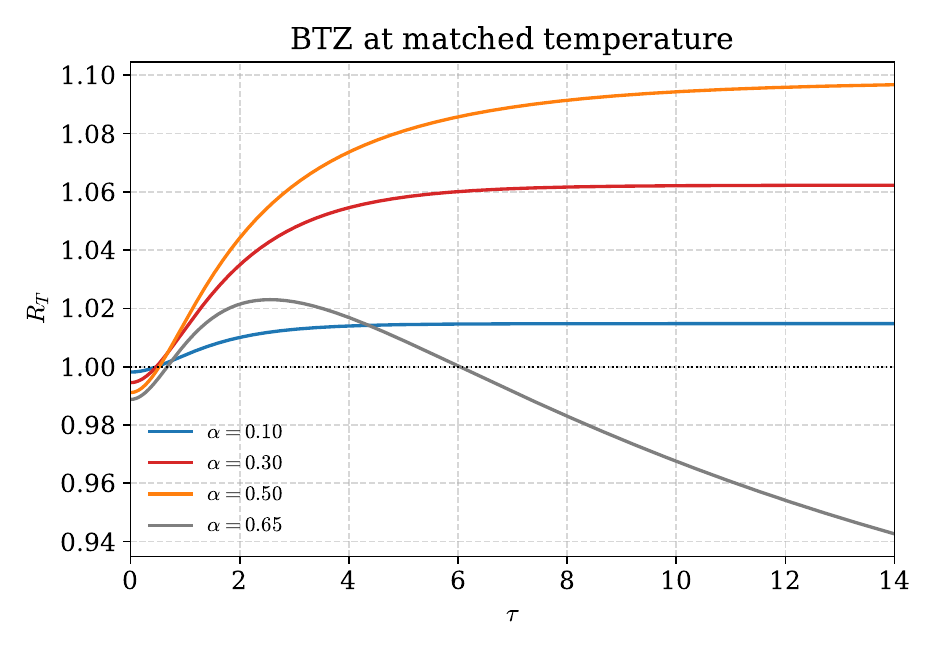}}
  \caption{\footnotesize{Accumulated geometric momentum in the Lorentzian-smeared background compared with ordinary BTZ at the same asymptotic mass (a) and at the same Hawking temperature (b). The physical release radius and probe normalization are kept common; the dotted line in (b) marks $R_T=1$. The ratios are evaluated only for $\tau>0$ because both numerator and denominator vanish at release.}}
  \label{fig:btzcomparison}
 \end{figure*}

\subsection{Gaussian mass profile}\label{subsec:gaussian}
To distinguish a feature of the momentum prescription from one tied exclusively to the Lorentzian tail, we also consider the illustrative dimensionless Gaussian-type lapse
\begin{equation}
 F^{\rm G}_\alpha(x)=x^2-1+\exp[-x^2/(4\alpha^2)].
 \label{eq:gaussian_lapse}
\end{equation}
This is a separate example; we do not assume that it follows from the same matter action as the Lorentzian solution. The same release-from-rest equations and normalization apply. At $X=3$, the early-time scale is $B_{\rm early}^{\rm G}\simeq1$ for $\alpha=0.2,0.3,0.45$, while the corresponding horizon scales are $\hat\kappa^{\rm G}=(0.98682,0.75522,0.12070)$. The qualitative separation between a weakly modified early response and a strongly modified near-horizon scale also occurs for this second profile. The Gaussian profile has $\alpha_c^{\rm G}=1/2$, equivalently $M_{\min}^{\rm G}=4\theta/l^2$. With $u=x^2$ and $s=4\alpha^2$, the lapse $h(u)=u-1+e^{-u/s}$ obeys $h(0)=0$ and $h''(u)>0$, with a positive outer root exactly when $s<1$. Setting $\epsilon_{\rm G}=1-4\alpha^2\to0^+$ yields
\begin{equation}
 x_+^{\rm G}\sim\sqrt{2\epsilon_{\rm G}},\qquad
 \hat\kappa_{\rm G}\sim\sqrt{2}\,\epsilon_{\rm G}^{3/2}.
\end{equation}
The critical exponents agree with the Lorentzian family but their coefficients differ. At fixed mass the Gaussian correction is exponentially small at large radius; consequently it does not change any coefficient of the algebraic Fefferman--Graham expansion, even though the horizon temperature and bulk geodesic lengths can change. The Gaussian mass profile corresponds to $M(r)=M[1-\exp(-r^2/(4\theta))]$, with the same definition $\alpha=\sqrt\theta/(l\sqrt M)$.

\begin{figure}[htbp]
\centering
\includegraphics[width=\columnwidth]{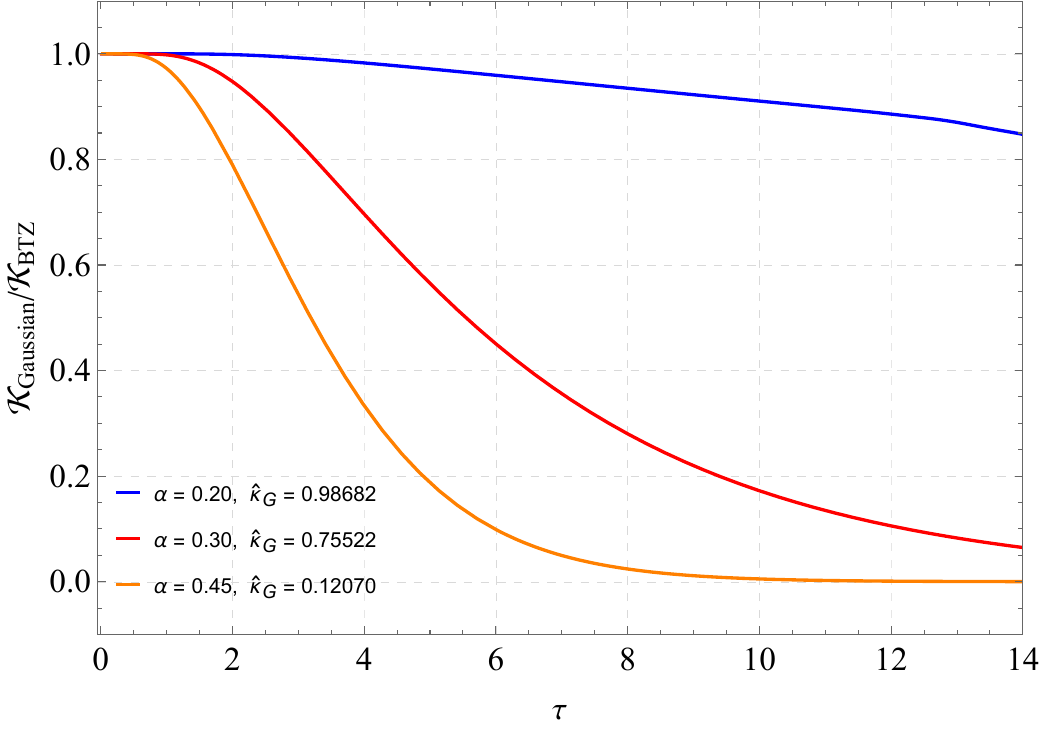}
\caption{\footnotesize{Finite-time response for the Gaussian-type lapse, normalized by the ordinary BTZ result at the same mass and release radius.}}
\label{fig:gaussian}
\end{figure}

The comparison is shown in Fig.~\ref{fig:gaussian}.

\section{Nonlocal geometric probes of Lorentzian smearing}\label{sec:nonlocal}
The Fefferman--Graham result in Sec.~\ref{sec:fg} raises a more discriminating question than the universal near-horizon momentum exponent: can an exterior, boundary-anchored observable detect the deformation at fixed asymptotic mass, even though $g_{(2)}$ is unchanged? We answer this question using equal-time spacelike geodesics.

For a connected equal-time geodesic with turning point $r_*>r_+$ and angular separation $\Delta\phi$, the conserved angular momentum of the length functional gives
\begin{align}
 \frac{\Delta\phi}{2}&=\int_{r_*}^{\infty}\frac{r_*\,dr}{r\sqrt{f_\theta(r)}\sqrt{r^2-r_*^2}},\label{eq:nonlocal_angle}\\
 L_\theta(R)&=2\int_{r_*}^{R}\frac{r\,dr}{\sqrt{f_\theta(r)}\sqrt{r^2-r_*^2}}.\label{eq:nonlocal_length}
\end{align}
The cutoff-independent difference at identical $M$, $\ell$, and boundary separation is $\Delta L=L_\theta^{\rm ren}-L_0^{\rm ren}$; the common ultraviolet logarithm cancels. For numerical stability, set $r=r_*/\sin u$, determine $r_*$ from Eq.~\eqref{eq:nonlocal_angle}, and evaluate the finite expression (here $\ell=M=1$)
\begin{equation}
\begin{split}
 L_\alpha^{\rm ren}={}&2\int_0^{\pi/2}\!\left[\frac{r_*}{\sin^2u\sqrt{F_\alpha(r_*/\sin u)}}-\frac{1}{\sin u}\right]du\\
 &-2\log r_*+2\log 2.
\end{split}
 \label{eq:nonlocal_finite}
\end{equation}
The comparison is made at the same asymptotic mass $M$, AdS scale $l$, angular interval, and radial cutoff $R$ before removing the cutoff. The finite additive constant is chosen so that the $\alpha=0$ result reproduces the standard BTZ expression $L_0^{\rm ren}=2\log[2\sinh(\Delta\phi/2)]$ for $M=\ell=1$; it cancels from $\Delta L$.

There is also an analytic monotonicity result. For $r>0$, $M>0$, and $q=\sqrt\theta>0$,
\begin{equation}
 \partial_q f_\theta(r)=\frac{Mr^2}{(r^2+q^2)^{3/2}}>0.
 \label{eq:nonlocal_monotone}
\end{equation}
On the common exterior region, increasing $q$ decreases the radial coefficient $g_{rr}=1/f_\theta$, leaving $g_{\phi\phi}=r^2$ unchanged. At fixed boundary endpoints, the length of each admissible spatial curve cannot increase. Because the outer horizon moves inward with increasing $q$, every exterior curve admissible for the smaller $q$ is also admissible for the larger $q$. Minimization therefore gives $L_{q_2}\le L_{q_1}$ for $q_2>q_1$ for the globally shortest exterior connected curve, and a strict inequality for nontrivial radial excursions. This statement concerns the connected exterior branch, not a complete competition among homology sectors on the boundary circle.

In the small-interval and weak-deformation regime, first-order variation of the geodesic functional on the undeformed BTZ curve yields, with $\ell=M=1$,
\begin{equation}
 \Delta L=-\frac{\pi}{128}\,\alpha(\Delta\phi)^3
 +O\!\left(\alpha(\Delta\phi)^5,\alpha^2\right).
 \label{eq:nonlocal_short}
\end{equation}
The cubic dependence reflects the $1/r$ asymptotic correction to $f_\theta$ and the corresponding $g_{(3)}$ coefficient, in contrast to the unchanged gravitational $g_{(2)}$. More generally, with the dimensionless interval $\Delta\phi$ and $q=\sqrt\theta$, the leading small-interval shift is $\Delta L=-\pi M q (\Delta\phi)^3/(128l)+O\!\left(\frac{q}{l}(\Delta\phi)^5,\frac{q^2}{l^2}(\Delta\phi)^3\right)$, where $L$ is measured in units of $l$; equivalently the physical length shift is $-\pi M q (\Delta\phi)^3/128$ at this order. The $g_{(3)}$ term is proportional to $Mq/l$ in the chosen FG coordinates, so the cubic shift is directly proportional to the leading smearing-dependent asymptotic datum. The scaling is compatible with a dimension-three boundary contribution; identifying a specific operator expectation value would require a complete bulk matter/dual-operator dictionary. The coefficient in Eq.~\eqref{eq:nonlocal_short} follows from the fixed-endpoint first variation; holding the turning point fixed would give an incorrect result.

\begin{figure}[t]
 \centering
 \includegraphics[width=\columnwidth]{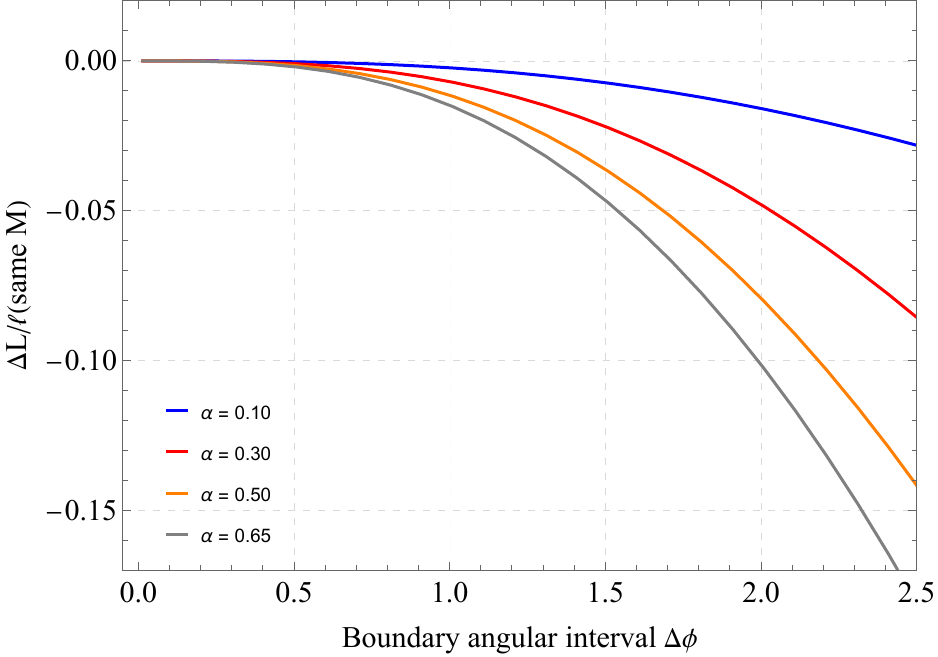}
 \caption{\footnotesize{Renormalized equal-time connected-geodesic length difference between Lorentzian-smeared and ordinary BTZ geometries at fixed $M=\ell=1$ and equal angular boundary separation. The full (unexpanded) metric is used. The negative shift grows in magnitude with smearing and interval size over the displayed range.}}
 \label{fig:nonlocal_length}
\end{figure}

Numerical quadrature confirms the monotonic length decrease for the tested intervals $\Delta\phi=0.5,1,2$ and smearings $\alpha=0.1,0.3,0.5,0.65$. For example, at $\Delta\phi=1$ the respective length shifts are $-0.0023134$, $-0.0069319$, $-0.0114882$, and $-0.0148295$. The finite additive normalization does not affect any length differences.

For the angular separations displayed in Fig.~\ref{fig:nonlocal_length}, we also compare the connected candidate to the complementary homology candidate,
\begin{equation}
 L_{\rm wrap}(\Delta\phi)=L_{\rm conn}(2\pi-\Delta\phi)+2\pi r_+,
 \label{eq:homology_competition}
\end{equation}
using the same renormalization prescription. Direct quadrature shows that the connected branch is shorter for $\Delta\phi\le 2.4$ at $\alpha=0,0.3,0.5,0.65$. At $\Delta\phi=2.4$ the respective connected/wrapping lengths are approximately $2.210/10.13$, $2.133/8.670$, $2.082/7.200$, and $2.047/5.357$ (for $l=M=1$). This verifies dominance within the plotted range, not for arbitrary intervals or all parameter values.

If a consistent holographic completion admits the standard Ryu--Takayanagi prescription for the relevant interval and connected homology branch, $\Delta S_A=\Delta L/(4G)<0$ \cite{Ryu:2006bv}. This interpretation is conditional, since the metric alone does not fix the matter action and its holographic boundary conditions. Independently of that interpretation, $\Delta L$ is a finite nonlocal geometric quantity that distinguishes the geometries at fixed asymptotic mass, complementing the local gravitational stress-tensor data and the proper-momentum response.

\subsection{Critical geometry and the momentum scale}\label{sec:critical}
The endpoint of the finite-radius horizon branch has a nontrivial scaling. With $\varepsilon=1-2\alpha^2\to0^+$, expansion of $F_\alpha(x)$ at the origin gives
\begin{equation}
 F_\alpha(x)=-\frac{\varepsilon}{2\alpha^2}x^2+\frac{3}{8\alpha^4}x^4+O(x^6),
 \qquad x_+\sim\frac{2\alpha}{\sqrt3}\varepsilon^{1/2},
 \label{eq:critical_horizon}
\end{equation}
so that
\begin{equation}
 \hat\kappa=\frac{F_\alpha'(x_+)}2\sim\frac{1}{\sqrt3\alpha}\varepsilon^{3/2},\qquad
 \hat\kappa^{-1}\propto\varepsilon^{-3/2}.
 \label{eq:critical_kappa}
\end{equation}
At fixed $q$ and $l$, the critical point corresponds to $M_{\min}=2q^2/l^2$, with $\varepsilon=1-M_{\min}/M$. Hence $\hat\kappa\propto(M-M_{\min})^{3/2}$ when approached from the black-hole branch. This zero-radius, zero-temperature limit is not by itself a proof of a dynamically stable remnant. The inverse surface gravity is a geometric near-horizon time, not a computed thermalization time of a boundary theory. The perturbative expansion at fixed $X>1$ is not uniform in the critical limit $\alpha\to1/\sqrt2$: the early-time coefficients remain regular there, whereas $B_{\rm late}\propto\varepsilon^{3/2}\to0$. This strengthens the scale separation on the nondegenerate branch but says nothing about a late-time exponential law at the endpoint itself. 

\section{Discussion and conclusions}\label{sec:discussion}

We have computed the proper radial momentum of a massive probe falling into a Lorentzian-smeared, nonrotating BTZ black hole and used it, through the momentum--Krylov prescription, as a complexity growth rate. The full Lorentzian mass profile was used throughout.

The geometry depends on the single parameter $\alpha=\sqrt\theta/(l\sqrt M)$. The outer horizon is
\begin{equation}
 x_+^2=1-\frac{\alpha^2}{2}-\frac{\alpha}{2}\sqrt{\alpha^2+4},
\end{equation}
and exists for $0\leq\alpha<1/\sqrt2$. As $\alpha$ increases, the horizon moves inward and the dimensionless surface gravity $\hat\kappa=F_\alpha'(x_+)/2$ decreases monotonically. The zero-radius degenerate root plays no role in the exterior infall.

In the BTZ limit the proper momentum is $|p_\rho|/m=X\sinh\tau$, which the numerical evolution reproduces. For nonzero smearing $|p_\rho|/m\propto e^{\hat\kappa\tau}$ at late times, and numerical fits agree with the analytic $\hat\kappa$ to about $1.3\times10^{-2}\%$. That the late-time exponent equals the surface gravity is a known property of the prescription \cite{Fan:2024iop,Caputa:2026}. Our results concern the full accumulated response $\mathcal K_\alpha(\tau)$. Its initial coefficient $F_\alpha'(X)/4$ and its late-time exponent $\hat\kappa$ depend differently on $\alpha$, so $\mathcal K_\alpha$ is neither a time-rescaled nor an amplitude- and time-rescaled BTZ curve; this follows analytically from $B_{\rm early}\neq B_{\rm late}$. For a distant release point the initial response approaches BTZ, while the late-time rate keeps its dependence on $\alpha$. The BTZ black hole with the same Hawking temperature and release radius has the same late-time exponent as the smeared geometry and reproduces its response to within about $10\%$; the remainder is the change of shape forced by $B_{\rm early}\neq B_{\rm late}$. The Gaussian profile behaves in the same way.

The surface gravity also fixes the Hawking temperature, so that $\lambda_{\rm geom}=2\pi k_BT_H/\hbar$. At fixed mass the smearing lowers both the growth rate of the proper momentum and the thermal scale of the MSS bound. This is a comparison of scales; no out-of-time-order correlator of the dual theory has been computed. Near $\alpha_c$ the quantities $\kappa$, $T_H$, and $\lambda_{\rm geom}$ vanish together along the nondegenerate branch, with critical exponents $1/2$ for the horizon radius and $3/2$ for the surface gravity.

The prescription~\eqref{eq:maincorrespondence} applies to a localized infalling probe and need not agree with other definitions of Krylov complexity in the boundary theory; in BTZ, operator and spread complexity are known to differ at late times \cite{Bhattacharya:2026quadratic,Bhattacharyya:2026interior}.

At fixed $M$, the Fefferman--Graham coefficient $g_{(2)}$, and hence the boundary stress tensor, are those of BTZ, while the smearing enters $g_{(3)}$. A shifted Hawking temperature leaves the stress tensor unchanged at fixed mass [Eq.~\eqref{eq:thermal_comparison}], which motivates observables beyond the stress tensor. The momentum--Krylov response and the connected geodesic lengths are two such observables: the first responds to the smearing through the horizon position and the surface gravity, and the second through the bulk geometry, where the connected lengths decrease. Assigning $g_{(3)}$ to a specific dual operator would require a complete holographic dictionary for the matter sector, and any entanglement interpretation of the geodesic lengths assumes a consistent matter-supported holographic completion.

Whether the smeared background is dual to a thermofield state at a shifted temperature or to a state with additional operator content \cite{Bak:2017} remains open. The nonconstant curvature, the matter source, and the restricted first-law identity show that temperature alone does not characterize the deformation. The nonlinear-electrodynamics parametrization is an effective reconstruction; its non-Maxwell behavior, noninvertible Legendre map, and unresolved parameter matching leave the dynamical completion and the microscopic dual for future work. None of these issues affects the classical probe calculation.

\acknowledgments
We would like to thank CNPq, CAPES and FAPESQ-PB. JAVC thanks the Paraíba State Research Support Foundation (FAPESQ) for financial support. F.A. Brito acknowledges CNPq (Grant no. 309092/2022-1) and Paraíba State Research Support Foundation (Universal-FAPESQ) for financial support.

\appendix
\section{Leading short-interval geodesic coefficient}
At fixed endpoints, the first-order length variation is evaluated on the undeformed BTZ geodesic. In the ultraviolet regime the Lorentzian correction begins at order $Mq/r$, while the leading turning point scales as $r_*\simeq 2l/\Delta\phi$ in the corresponding dimensionless boundary coordinates. The relevant dimensionless integral reduces to
\begin{equation}
 \int_1^\infty \frac{\sqrt{s^2-1}}{s^5}\,ds=\frac{\pi}{16}.
\end{equation}
Combining the fixed-endpoint variation with the turning-point scaling yields the cubic correction quoted in Eq.~(\ref{eq:nonlocal_short}); the correction must be expressed in powers of $q/l$ when lengths are normalized by $l$.

\section*{Data and code availability}
The numerical data and the Python scripts used to generate the figures and tables are available from the authors upon reasonable request.


\begin{thebibliography}{99}

\bibitem{Carlip:1995qv} 
  S.~Carlip, The (2+1)-Dimensional black hole, Class.\ Quant.\ Grav.\  {\bf 12}, 2853 (1995).

\bibitem{AdS} J. M. Maldacena, The Large N Limit of Superconformal Field Theories and Supergravity, Adv. Theor. Math. Phys. {\bf 2}, 231 (1998).

\bibitem{BTZ} M. Banados, C. Teitelboim, and J. Zanelli, The Black Hole in Three Dimensional Space Time, Phys. Rev. Lett. {\bf 69}, 1849 (1992).

\bibitem{Chan:1995wj} K.~C.~K.~Chan and R.~B.~Mann, Spinning black holes in (2+1)-dimensional string and dilaton gravity, Phys.\ Lett.\ B {\bf 371}, 199 (1996).

\bibitem{Martinez:1996gn} C.~Martinez and J.~Zanelli, Conformally dressed black hole in (2+1)-dimensions, Phys.\ Rev.\ D {\bf 54}, 3830 (1996).

\bibitem{Chan:1996rd} K.~C.~K.~Chan, Modifications of the BTZ black hole by a dilaton / scalar, Phys.\ Rev.\ D {\bf 55}, 3564 (1997).

\bibitem{Banados:2001}M. Bañados, O. Chandía, N. E. Grandi, F. A. Schaposnik, and G. A. Silva, Chern-Simons formulation of noncommutative gravity in three dimensions, Physical Review D, vol. 64, no. 8, article 084012, 2001.

\bibitem{Kim:2008}H.-C. Kim, M.-I. Park, C. Rim, and J. H. Yee, Smeared BTZ black hole from space noncommutativity, Journal of High Energy Physics, vol. 10, p. 60, 2008.

\bibitem{Chang-Young:2009}E. Chang-Young, D. Lee, and Y. Lee, The noncommutative BTZ black hole in polar coordinates, Classical and Quantum Gravity, vol. 26, article 185001, 2009.

\bibitem{Jahangir:2011}R. Jahangir and K. Saifullah, Thermodynamics of noncommutative BTZ black hole, https://arxiv.org/abs/1101.6073.

\bibitem{Liang:2012}J. Liang and B. Liu, Thermodynamics of noncommutative geometry inspired BTZ black hole based on Lorentzian smeared mass distribution, EPL, vol. 100, no. 3, p. 30001, 2012.

\bibitem{Nicolini:2005vd} 
  P.~Nicolini, A.~Smailagic and E.~Spallucci, Noncommutative geometry inspired Schwarzschild black hole, Phys.\ Lett.\ B {\bf 632}, 547 (2006).

\bibitem{Nicolini:2008aj} P.~Nicolini, Noncommutative Black Holes, The Final Appeal To Quantum Gravity: A Review, Int.\ J.\ Mod.\ Phys.\ A {\bf 24}, 1229 (2009).

\bibitem{Nozari:2008rc} 
  K.~Nozari and S.~H.~Mehdipour, Hawking Radiation as Quantum Tunneling from Noncommutative Schwarzschild Black Hole,  Class.\ Quant.\ Grav.\  {\bf 25}, 175015 (2008).

\bibitem{Anacleto:2014cga} M.~A.~Anacleto, F.~A.~Brito and E.~Passos, Gravitational Aharonov - Bohm effect due to noncommutative BTZ black hole, Phys.\ Lett.\ B {\bf 743}, 184 (2015).

\bibitem{Anacleto:2019tdj} 
  M.~A.~Anacleto, F.~A.~Brito, J.~A.~V.~Campos and E.~Passos, Absorption and scattering of a noncommutative black hole, Phys.\ Lett.\ B {\bf 803}, 135334 (2020).

\bibitem{Rahaman:2013gw} 
  F.~Rahaman, P.~K.~F.~Kuhfittig, B.~C.~Bhui, M.~Rahaman, S.~Ray and U.~F.~Mondal, BTZ black holes inspired by noncommutative geometry, Phys.\ Rev.\ D {\bf 87}, no. 8, 084014 (2013).

\bibitem{Sadeghi:2015nzp} 
  J.~Sadeghi and V.~R.~Shajiee, Effective Potential in Noncommutative BTZ Black Hole, Int.\ J.\ Theor.\ Phys.\  {\bf 55}, no. 2, 892 (2016).

\bibitem{Anacleto:2015kca} 
  M.~A.~Anacleto, F.~A.~Brito, A.~G.~Cavalcanti, E.~Passos and J.~Spinelly, Quantum correction to the entropy of noncommutative BTZ black hole, Gen.\ Rel.\ Grav.\  {\bf 50}, no. 2, 23 (2018).

\bibitem{Kawamoto:2017ufe} 
  S.~Kawamoto, K.~Nagasaki and W.~Y.~Wen, Charged rotating BTZ black holes in noncommutative spaces and torsion gravity, PTEP {\bf 2018}, no. 4, 043E01 (2018).

\bibitem{Nandy:2024evd}
P. Nandy, A. S. Matsoukas-Roubeas, P. Martínez-Azcona, A. Dymarsky and A. del Campo, Quantum dynamics in Krylov space: Methods and applications, Phys. Rep. \textbf{1125--1128}, 1--82 (2025), doi:10.1016/j.physrep.2025.05.001.

\bibitem{Baiguera:2025dkc} S. Baiguera, V. Balasubramanian, P. Caputa, S. Chapman, J. Haferkamp, M. P. Heller et al.,
Quantum complexity in gravity, quantum field theory, and quantum information science, Phys. Rep. \textbf{1159}, 1--77 (2026), doi:10.1016/j.physrep.2025.11.001, arXiv:2503.10753 [hep-th].

\bibitem{Rabinovici:2025otw} E. Rabinovici, A. Sánchez-Garrido, R. Shir and J. Sonner, Krylov Complexity, 2507.06286.

\bibitem{Douglas:2014}Douglas Stanford and Leonard Susskind, Complexity and Shock Wave Geometries,  Phys. Rev. D 90, 126007 (2014).

\bibitem{Adam:2016} A. R. Brown et al., Complexity, action, and black holes, Phys.Rev.D 93 (2016) 8, 086006.

\bibitem{Brown:2016}A.R. Brown, D.A. Roberts, L. Susskind, B. Swingle, Y. Zhao, Holographic complexity equals bulk action? Phys. Rev. Lett. 116, 191301 (2016).

\bibitem{Chapman:2017}S. Chapman, H. Marrochio, R.C. Myers, Complexity of formation in holography. JHEP 01, 062 (2017).

\bibitem{Caputa:2026} P. Caputa, B. Chen, R. W. McDonald, J. Simón, and B. Strittmatter, Spread complexity rate as proper momentum, Phys. Rev. D 113, L041901 (2026), doi:10.1103/7zs8-9zpg, arXiv:2410.23334 [hep-th].

\bibitem{Fan:2024iop}Z.-Y. Fan, Momentum-Krylov complexity correspondence, arXiv:2411.04492 [hep-th] (2024).

\bibitem{He:2024} P.-Z. He, Revisit the relationship between spread complexity rate and radial momentum, arXiv:2411.19172 [hep-th] (2024).

\bibitem{Alfinito:2026} E. Alfinito and M. Beccaria, Krylov correlators in $\mathfrak{sl}(2,\mathbb R)$ models: exact results and holographic complexity, arXiv:2605.17550 [hep-th] (2026).

\bibitem{Fatemiabhari:2026}A. Fatemiabhari and C. N\~unez, Krylov complexity, confinement and universality, arXiv:2602.17757 [hep-th] (2026).

\bibitem{Bhattacharya:2026quadratic}
A.~Bhattacharya, M.~Flory, M.~P.~Heller, E.~Rizza, and T.~Schuhmann,
The quadratic growth of Krylov spread complexity in the BTZ black hole,
arXiv:2608.09922 [hep-th] (2026).

\bibitem{Bhattacharyya:2026interior}
A.~Bhattacharyya, S.~Pal, and J.~F.~Pedraza,
Krylov complexity and the growth of the black hole interior in 3D gravity,
arXiv:2608.19373 [hep-th] (2026).

\bibitem{Maldacena:2015waa}
J.~Maldacena, S.~H.~Shenker, and D.~Stanford,
A bound on chaos,
JHEP \textbf{08}, 106 (2016), arXiv:1503.01409 [hep-th].

\bibitem{Avdoshkin:2022xuw}
A.~Avdoshkin, A.~Dymarsky, and M.~Smolkin,
Krylov complexity in quantum field theory, and beyond,
JHEP \textbf{06}, 066 (2024), arXiv:2212.14429 [hep-th].

\bibitem{Hashimoto:2023swv}
K.~Hashimoto, K.~Murata, N.~Tanahashi, and R.~Watanabe,
Krylov complexity and chaos in quantum mechanics,
JHEP \textbf{11}, 040 (2023), arXiv:2305.16669 [hep-th].

\bibitem{Bak:2017} D.~Bak et al., ``Holographic Micro Thermofield Geometries of BTZ Black Holes,'' JHEP \textbf{06}, 079 (2017), arXiv:1704.01030.

\bibitem{BrownHenneaux:1986} J.~D.~Brown and M.~Henneaux, ``Central Charges in the Canonical Realization of Asymptotic Symmetries: An Example from Three-Dimensional Gravity,'' Commun. Math. Phys. \textbf{104}, 207 (1986).

\bibitem{deHaro:2000} S.~de Haro, S.~N.~Solodukhin, and K.~Skenderis, ``Holographic reconstruction of spacetime and renormalization in the AdS/CFT correspondence," Commun. Math. Phys. \textbf{217}, 595 (2001), arXiv:hep-th/0002230.

\bibitem{Ryu:2006bv} S.~Ryu and T.~Takayanagi, ``Holographic derivation of entanglement entropy from AdS/CFT," Phys. Rev. Lett. \textbf{96}, 181602 (2006), arXiv:hep-th/0603001.

\end{thebibliography}
\end{document}